\documentclass[sn-mathphys-ay]{sn-jnl}
\usepackage{setspace}
 \usepackage[switch]{lineno}
\usepackage{graphicx}%
\usepackage{multirow}%
\usepackage{amsmath,amssymb,amsfonts}%
\usepackage{mathrsfs}%
\usepackage[title]{appendix}%
\usepackage{xcolor}%
\usepackage{textcomp}%
\usepackage{manyfoot}%
\usepackage{booktabs}%
\usepackage{algorithm}%
\usepackage{algorithmicx}%
\usepackage{algpseudocode}%
\usepackage{listings}%
\usepackage{arydshln}

\chardef\us=`\_

\begin{document}

\title[]{Dynamics and solar wind control of the recovery of strong geomagnetic storms}


\author[1,2]{\fnm{O.} \sur{Ahmed}}\email{osmanr18@ymail.com;oahmedindris@stu.kau.edu.sa}

\author*[1]{\fnm{B.} \sur{Badruddin}}\email{badr.physamu@gmail.com; bzahmad@kau.edu.sa}
\equalcont{These authors contributed equally to this work.}

\author[1]{\fnm{M.} \sur{Derouich}}\email{derouichmoncef@gmail.com}
\equalcont{These authors contributed equally to this work.}

\affil*[1]{\orgdiv{Astronomy and Space Science Department}, \orgname{Faculty of Science, King Abdulaziz University}, \orgaddress{\street{Jeddah 21589}, \city{Jeddah}, \postcode{P.O. Box 80203}, \state{Makkah}, \country{Saudi Arabia}}}

\affil[2]{\orgdiv{Department of Physics}, \orgname{Faculty of Natural and computational Sciences, Debre Tabor University}, \orgaddress{\street{Street}, \city{Debre Tabor}, \postcode{P.O. Box 272}, \state{South Gondar}, \country{Ethiopia}}}

\abstract{In this work we have studied about the characteristics and dynamical changes during the recovery time of moderate and strong geomagnetic storms of ($\mathrm{Dst}<-50$ nT).
In our investigation of 57 storms triggered by CMEs/CIRs, we concentrated on the solar wind's influence on their decay phases. Selected storms were classified into distinct groups based on their recovery characteristics. 
Employing the superposed epoch analysis and best fit methods, we scrutinized several interplanetary solar wind plasma and field parameters and their various functions. The analysis encompassed various single, dual, and multiple interplanetary plasma and field parameters/functions.
We determined the most representative characteristic time for the storm's recovery profile by carefully fitting an exponential curve. A correlation analysis between Dst and solar wind parameters/functions led us to isolate a coupling function ($\rho^{\frac{1}{2}}$Ey) which best described the decay rate of the ring current. It shows that electric field term (Ey) coupled with a viscus term ($\rho^{\frac{1}{2}}$) plays pivotal role in determining the recovery rate of a geomagnetic storms.
Additionally, we modeled the complex patterns of Dst recovery in relation to solar wind parameters and functions using a second-order polynomial.
Remarkably, during the recovery phase, a dynamic correlation between Dst and solar wind parameters/functions was revealed. The three-parameter solar wind-magnetosphere electrodynamical coupling functions, which combines the viscus term ($\rho^{\frac{1}{2}}$) and the electric field-related function (v$^{\frac{4}{3}}$B) ($\rho^{\frac{1}{2}}$v$^{\frac{4}{3}}$B), significantly impacts the recovery phase of geomagnetic disturbances.
Our investigation extended to the relationship between main and recovery phase durations, providing valuable insights into the solar wind's intricate control over the decay of the geomagnetic disturbances. These findings contribute significantly to advancing our comprehension of the complex relationship between solar wind dynamics and the evolution of geomagnetic disturbances.}




\maketitle

\section{Introduction}\label{sec:intro} 
Coronal mass ejections (CMEs) in general are found to be the sources of major geomagnetic storms (GSs) \citep[e.g.,][]{2012LRSP....9....3W}, while corotating interaction regions (CIRs) are more commonly responsible for moderate and weak magnetic storms. However, Bz component of the interplanetary magnetic field (IMF) typically fluctuates greatly within CIRs results weak storms \citep[e.g.,][]{1999SSRv...88..529G, 2006ilws.conf..266G, 2009JASTP..71..885G} 
that require more time to recover than CMEs \citep[e.g., see][and references therein]{2006JGRA..111.1103X, 2023AdSpR..71.5147P}. These phenomena are extensively studied \citep[e.g.,][]{2008SoPh..250..375J, 2008JASTP..70.2078G, 2021Univ....7..138Y}. 
In recent years, the driver sources of GSs have been identified in greater detail \citep[e.g.,][]{2007JGRA..11210102Z, 2008JGRA..113.5221E, 2012JSWSC...2A..01R, 2021Univ....7..138Y}. \\
Much work has been done in recent years to better understand the causes behind magnetosphere turbulence and to make better predictions before transients, primarily from the Sun threaten the magnetosphere. 
GSs mostly develop in two clearly distinct phases; main and recovery. The majority of previous research concentrated on the storm's main phase \citep[e.g.,][]{1994JGR....99.5771G, 2016Ap&SS.361..253B, 2023SpWea..2103314M, 2024arXiv240203261A}.
However, relatively fewer studies were conducted to gain a thorough understanding of the post-main phase GS activity known as the storm recovery phase \citep[e.g.,][]{YACOB_1964, 2012JGRA..117.8207Y, 2014LNEA....3..127A, 2021JGRA..12628685C}.\\
Geomagnetic indices were introduced decade ago \citep[e.g.,][]{Mayaud1980DerivationMA, Saba1997RelationshipsBT}, and now a days scholars are using them to investigate extraterrestrial interactions with the Earth's magnetosphere, and among them Dst is the most studied geomagnetic index \citep[e.g.,][]{2021EJGeo...2....1L, 2022JPhCS2214a2024N, 2023SpWea..2103304B, 2023Atmos..14.1744M, 2024arXiv240203261A}.
Previous studies show solar wind-magnetosphere interactions \cite[e.g.,][]{Badruddin1998InterplanetarySM, Kane2007PhaseS, 2008JGRA..113.5221E, Khabarova2008SolarWP, Badruddin2009GeoeffectivenessOM} which are the the main drivers for GS activity \citep[e.g.,][]{Gosling1999FormationAE, Wang2002ASS, 2008JASTP..70.2078G}. The relationship between GS and interplanetary (IP) plasma and field parameters solar wind (SW) speed v, plasma density $\rho$ and IMF Bz have been studied and suggested to be strongly related with the intensity of severe GSs \cite[e.g.,][]{Garrett1974InfluenceOS, 2008PhyS...78d5902M}.
The horizontal component of low latitude magnetic fields has the
magnitude reduced during one to several hours during the main phase of GSs, and it can take several days to recover \citep[e.g.,][]{Lakhina2018}.
In addition to the single plasma and field parameters, derived solar wind-magnetosphere electrodynamical CFs have also been studied \cite[e.g.,][]{Srivastava2004SolarAI, Kane2005HowGI, Singh2006AnalysisOP, 2023SpWea..2103314M} and many of them reported that IP electric field, during southward IMF is best correlated with the geomagnetic index Dst. \\ 
Long-lasting southward IMF Bz is the primary source of severe magnetic storms \citep[e.g.,][]{2008JGRA..113.5221E} which increase the magnetic reconnection of energy injection into the magnetosphere \citep[e.g.,][]{1997GMS....98...77T} and the ring current enhancement results from this plasma injection \citep[e.g.,][]{Lakhina2018}. 
It is made up of electrons and ions with energies ranging from about 1 -- 400 KeV, primarily H$^+$, O$^+$ \citep[e.g.,][]{Vallat2005FirstCD, Liemohn2016ChallengesAW}. 
Strong GSs resulted from a high SW dynamic pressure amplifying the energy of the ring current in the presence of a consistent southern IMF Bz \citep[e.g.,][]{2010JASTP..72.1364R}. 
Ionospheric disturbances are more significant during the GSs recovery phase than before its onset \citep[e.g.,][]{2016AdSpR..58.2057C}. Recent work by \cite{yan2022ring} suggested that geomagnetic activity is closely related to the role that O$^+$ ions play in ring current. However, H$^+$ is the predominant ion in the plasma sheet during the quiet-to-disturbance transition \citep[e.g.,][]{2017JGRA..12212040D}.
\\
A number of studies have previously examined that the O$^+$ charge exchange loss rate concurrent with Dst dominates the faster first step recovery from the peak of Dst while the H$^+$ charge exchange loss rate controls the slower later step recovery \citep[e.g.,][]{1988JGR....9314343H, 1997GMS....98..107D} the  superposition and resultant loss rates of O$^+$ loss rates are roughly linear \citep[e.g.,][]{2014PhPl...21c2903J}. 
\cite{1999GeoRL..26.2845L} reported that convective drift loss out of the day side magneto-pause was found to be the primary mechanism removing ring current particles during the initial recovery of GSs.
The flow out effect and the charge exchange-induced loss of confined particles are superposed to explain the recovery of the ring current \citep[e.g.,][]{1990P&SS...38.1133T}. 
\cite{2009JASTP..71..885G} pointed out that, after the IP event, the high-speed stream's presence seems to have an impact on the storm's recovery.
In addition to the constant loss of ring current particles due to common physical processes, magnetic reconnection IMF Bz Alfven waves and magnetopause fields are also constantly injecting new particles into the magnetosphere's outer regions for CIR events \citep[e.g.,][]{2006JGRA..111.7S01T}. Later \cite{2018ApJ...860...26R} confirm that, Alfven waves are responsible for the lengthy recovery period of GSs by introducing energy into the magnetosphere during the storm. \cite{2019MNRAS.488..910R} discussed that, not only weak/moderate (CIR driven) storms recovery is extended by Alfven waves but also strong storms. 
\cite{2019MNRAS.490.3440S} suggested that Alfven waves may contribute to slow recovery of storms. Physical processes such as recombination, in particular, are responsible for the sluggish recovery of alfven waves \citep[e.g.,][]{Telloni_2021}.
According to theoretical simulations, it was observed that the injection of the ring current is influenced by both the dynamic pressure and electric field of the SW \citep[e.g.,][]{2022JGRA..12730404G}.\\
Most of previous studies primarily examined individual events, limited to a particular solar cycle or specific GS intensities. However, our study takes a different approach by investigating the recovery phase time properties and dynamics of GSs across three solar cycles encompassing the years 1995 to 2022. Additionally, we analyze averaged storms originating from diverse drivers and varying intensities, by employing the superposed epoch analysis.\\
Convective electric field (-vBz) effects on recovery time were first taken into account by \citep[e.g.,][]{1998GeoRL..25.2999F}. 
The suggestion of such CFs during the recovery phase motivates us to introduce more electric field-related functions, which are later combined with (viscus term $\rho$ and dynamic pressure term P) \citep[e.g.,][]{2008JGRA..113.4218N} with different power indices ($\frac{1}{2}$, $\frac{1}{3}$). We obtained a useful representative derived function from such combination that expressed the recovery time characteristics better than individual component alone. Electric field functions paired with a viscus term best reflect the recovery time features of GSs, \citep[also see,][]{OBrien2000AnEP}.
Exponential constant ($\tau$), which varies over time and is dependent on the minimum Dst, can be used to approximate the Dst profile \citep[e.g.,][]{OBrien2000AnEP, 2008AnGeo..26.2543M, 2012JGRA..117.8207Y}. While anti-correlation exists for long duration recovery, (Dst)$_\mathrm{peak}$ intensity is correlated with short durations \citep[e.g.,][]{Aguado2010AHD, 2012JGRA..117.8207Y, Cid2014OnEG}.
\cite{2021Ap&SS.366...62B} recently applied exponential fitting to the rate of decay time profile of Forbush decrease events. To describe the rate of Dst decay during the recovery period, we used an exponential curve fitting model.\\
In the present study, the recovery features of GSs have been investigated using the geomagnetic index Dst, as well as numerous IP SW parameters and their derived functions. Superposed epoch analysis and correlation analysis methods were employed, focusing on the recovery phase of geomagnetic disturbances. This study attempts to explore the fundamental dynamics and time aspects of the GS recovery profile. The goal is to identify the optimal IP SW parameter/function for controlling ring current decay during GSs, taking into account the Dst recovery time constant and morphology.
Data retrieval, classification, and analysis techniques is covered in Section \ref{sec:data}. In Section \ref{sec:res}, the results based on superposed analysis subjected to exponential curve fitting, and the relationship of the geomagnetic index Dst with the other IP plasma and field parameters and their derivatives based on morphology and characteristic recovery time were explored. The results of our effort has been summarized and explained in section \ref{sec:conc}.

\section{Observations and data analysis}\label{sec:data}

In this study, we identified 57 GS events Dst ($\mathrm{Dst}\leq-50$ nT) with specified criteria (mentioned below) measured with 1-hour resolution data from 1995 to 2022, spanning three solar cycles.
We used the NSSDC OMNI\footnote{\url {https://omniweb.gsfc.nasa.gov/ow.html}} database to retrieve averaged hourly geomagnetic index Dst, IP plasma, and field parameters.  
The storms that satisfied the following criteria were chosen:
\begin{itemize}
	\item Although the peak intensity and duration of recovery vary among geomagnetic storm events, the recovery pattern should ideally follow an exponential curve, sharp recoveries over short time spans and extend recovery beyond the onset level are not considered.
	\item The smoothness of the recovery profiles is taken into account, and some zigzag-like dip structures with high amplitudes are excluded.
	\item Full or nearly full recovery profiles with different recovery time are considered
\end{itemize}
The majority of the chosen GS events are caused by IP CMEs, as each event was recognized from near-Earth IP  CMEs{\footnote{\url{https://izw1.caltech.edu/ACE/ASC/DATA/level3/icmetable2.htm}}} \citep[e.g.,][]{2003JGRA..108.1156C}.\\
In order to isolate CFs that best represents the decay of the ring current during the recovery phase of GSs, we have considered  geomagnetic index Dst, IP SW parameters and their derived functions. 
We have retrieved two hourly averaged IMF parameters namely (IMF magnitude B (nT), the north-south component of IMF Bz (nT), and related variance represented by sigma field vector ($\sigma$B (nT)), four IP plasma parameters (proton density $\rho$ (cm$^{-3}$), SW speed v (kms$^{-1}$), pressure P (nPa) and dawn-dusk electric field Ey (mVm$^{-1}$) and multiple product CFs
derived from combinations of viscus terms ($\rho$ and P) coupled with electric field related functions (e.g., Ey \& v$^{\frac{4}{3}}$Bz) v and IMF variance related functions ($\sigma$B) 
[v$\sigma$B, v$^{\frac{4}{3}}$Bz, $\rho$Ey, $\rho^{\frac{1}{2}}$Ey, $\rho^{\frac{1}{3}}$Ey, PEy, P$^{\frac{1}{2}}$Ey, P$^{\frac{1}{3}}$Ey, $\rho$v$\sigma$B, $\rho^{\frac{1}{2}}$v$\sigma$B, $\rho^{\frac{1}{3}}$v$\sigma$B, Pv$\sigma$B, P$^{\frac{1}{2}}$v$\sigma$B, P$^{\frac{1}{3}}$v$\sigma$B, $ \rho$v$^{\frac{4}{3}}$Bz, $\rho^{\frac{1}{2}}$v$^{\frac{4}{3}}$Bz, $\rho^{\frac{1}{3}}$v$^{\frac{4}{3}}$Bz, Pv$^{\frac{4}{3}}$Bz, P$^{\frac{1}{2}}$v$^{\frac{4}{3}}$Bz, P$^{\frac{1}{3}}$v$^{\frac{4}{3}}$Bz and $\rho^{\frac{1}{2}}$v$^{\frac{4}{3}}$B] as explained in Section \ref{sec:res}.   \\
Selection of these functions, was motivated by earlier suggested CFs representing geomagnetic disturbances and possible mechanisms operating during the ring current development and decay \citep[e.g., see,][]{1981P&SS...29.1151A, 1989JGR....94.8835G, 1998GeoRL..25.2999F, 1999GeoRL..26.2845L, 2006JGRA..111.7S01T, 2007JGRA..112.1206N, 2008JGRA..113.4218N, 2023AdSpR..71.1137B, 2024arXiv240203261A}. Some new variants of earlier explored CFs have also been tested.
We chose geomagnetic index, IP plasma, and field parameters, together with their product functions, for their potential importance to storm recovery. 
To model the storm's recovery profile, we divided the data into twelve groups based on the main and recovery phases of the time profiles and fitted them with the relevant equations. The source-based classification is also provided. We employed exponential curve fitting to fully understand recovery time patterns, which is described in detail on later Section \ref{curve}.
To fit our data, we used the exponential equation: $\mathrm{Dst(x)} = \mathrm{(Dst)_{min}}+\mathrm{Ae^{-(x-x_0)/\tau}}$,
 where $\tau$ is the time constant, $\mathrm{(Dst)_{min}}$ is its peak value.
 Our focus is on the time constant $\tau$, which provides information on the rate of the storm recovery tabulated later in Table \ref{t1}. 
Earlier studies employed the exponential decay function to represent the fast recovery phase of the storm \citep[e.g., see,][and references therein]{Aguado2010AHD, 2021JGRA..12628685C}.
Most of earlier studies applied the exponential fit procedure to the recovery of a few selected individual storm events. However, in most cases the progress of recovery is highly fluctuating in its nature.
\\
To facilitate our analysis and ensure a smoother recovery profile, we divided the selected storms into groups and analyzed averaged data.
Four groups are examined based on recovery period, which is measured from the storm's peak value (Dst minimum) until its return to normal levels.
The first group, labeled as r$^1$, represents \textquotedblleft very fast\textquotedblright recovery with a recovery time of $\leq 4$ days and consists of 16 events. The second group, labeled as r$^2$, denotes \textquotedblleft fast\textquotedblright recovery, taking 4 -- 6 days to reach the onset level, and comprises 14 events. The third group, labeled as r$^3$, represents \textquotedblleft slow\textquotedblright recovery, requiring 6 -- 9 days to reach onset level, and consists of 13 events. The remaining 14 events are classified as r$^4$, representing \textquotedblleft very slow\textquotedblright recovery, taking more than 9 days to return to quiet level. The source-based classification of two groups and one special group, which accounts for intense storms (Dst $\leq-150$ nT), is also indicated in Table \ref{tabl1}.
To get a comprehensive conclusion, the remaining five main phase-based groupings are also examined and defined in the table notes of Table \ref{t1}.\\
We obtained the average recovery profile of Dst for each group, as well as the concurrent profiles of several IP SW parameters/functions.
Using the results of curve fit time constants, we examined the relationship between Dst and IP SW parameters/functions.
By investigating the correlations, we gained insights into the parameters/functions that are more closely related to the temporal profiles of storm recovery.
Furthermore, we explored the association between Dst and IP SW parameters/functions based on recovery morphology.
By evaluating the correlation, we acquired insight into the parameters that are more related to the recovery dynamics of GSs.

\begin{table}
\caption{Log of the selected GS data with Dst recovery time profiles and classification of storms based on criteria described on Section \ref{sec:data}. }\label{tabl1}
\begin{tabular}{@{}llllll@{}}
\toprule%
No  &  Recovery Start &Recovery &(Dst)min&Recovery&Sources\\
&date \& time(UT)&time(hrs)$^*$&(nT)&group&\\
\midrule
                1  &1995-04-07T19:00&247& -149& r$^4$&CIR\\
                2  &1995-10-19T00:00&22& -127&  r$^2$&CME\\
				3  &1997-01-10T10:00&16 & -78&  r$^1$&CME\\
				4  &1997-04-22T00:00&72& -107& r$^3$&CME\\
				5  &1997-05-02T01:00&196& -64& r$^2$&CIR\\
				6  &1997-05-15T13:00&262& -115& r$^2$&CME\\
				7  &1997-05-27T07:00&243& -73& r$^3$&CME\\
				8  &1997-06-09T05:00&50& -84& r$^3$&CME\\
				9  &1997-09-03T23:00&109& -98& r$^4$&CME\\
				10 &1997-10-11T04:00&94& -130& r$^2$&CME\\
				11 &1997-11-07T05:00&150 & -110& r$^1$&CME\\
				12 &1998-01-07T05:00&30 & -77& r$^3$&CME\\
				13 &1998-03-10T21:00&218 & -116& r$^2$&CIR\\
				14 &1998-06-26T05:00&132 & -101& r$^2$&CME\\
				15 &1998-08-06T12:00&187 & -138& r$^2$&CME\\
				16 &1998-08-27T10:00&204 & -155$^\Lambda$& r$^4$&CME\\
				17 &1998-09-25T08:00&120 & -202$^\Lambda$& r$^1$&CME\\
				18 &1998-11-13T22:00&108 & -131& r$^3$&CME\\
				19 &1999-01-14T00:00&109 & -112&r$^3$&CME\\ 
				20 &1999-09-23T00:00&92 & -173$^\Lambda$& r$^1$&CME\\  
				21 &2000-01-11T23:00&177 & -80& r$^3$&CIR\\
				22 &2000-04-07T01:00&138 & -292$^\Lambda$& r$^1$&CME\\ 
				23 &2000-07-16T01:00&90 & -300$^\Lambda$& r$^2$&CME\\
				24 &2000-08-12T10:00&60 & -234$^\Lambda$& r$^1$&CME\\
				25 &2000-09-18T00:00&162 & -201$^\Lambda$& r$^1$&CME\\ 
				26 &2001-03-31T09:00&103 & -387$^\Lambda$& r$^1$&CME\\
				27 &2001-04-12T00:00&35 & -271$^\Lambda$& r$^1$&CME\\
				28 &2001-04-18T02:00&61 & -114& r$^1$&CME\\
				29 &2001-04-22T16:00&134 & -102& r$^3$&CME\\
				30 &2001-08-17T22:00&89 & -105& r$^2$&CME\\
				31 &2001-10-03T15:00&34 & -166$^\Lambda$& r$^1$&CME\\
				32 &2001-11-06T07:00&226 & -292$^\Lambda$& r$^2$&CME\\
				33 &2001-11-24T17:00&149 & -221$^\Lambda$&r$^2$&CME\\
                34 &2002-09-08T01:00&147 & -181$^\Lambda$& r$^2$&CME\\
				35 &2003-11-20T22:00&170 & -422$^\Lambda$& r$^4$&CME\\
				36 &2004-04-04T01:00&32 & -117& r$^2$&CME\\
				37 &2004-08-30T23:00&124 & -129& r$^3$&CME\\
				38 &2005-01-22T07:00&133 & -97& r$^2$&CME\\
				39 &2005-05-15T09:00&311 & -247$^\Lambda$& r$^1$&CME\\
				40 &2006-12-15T08:00&85 & -162$^\Lambda$& r$^3$&CME\\
				41 &2009-07-22T07:00&198  &-83& r$^1$&CME\\ 
				42 &2011-08-06T04:00&198  &-115& r$^1$&CME\\
				43 &2011-10-24T18:00&129 & -147& r$^1$&CME\\
				44 &2012-04-24T05:00&285 & -120& r$^4$&CME\\
				45 &2012-07-15T19:00&283 & -139& r$^3$&CME\\

\botrule
\end{tabular}
\end{table}
\begin{table}[h]
\begin{tabular}{@{}llllll@{}}
\toprule%
No  &  Recovery Start &Recovery &(Dst)min&Recovery&Sources\\
&date \& time(UT)&time(hrs)\textbf{$^*$}&(nT)&group&\\
\midrule
				46 &2012-11-14T08:00&231 & -108& r$^4$&CME\\
				47 &2013-03-17T21:00&221 & -132& r$^3$&CME\\
				48 &2013-06-01T09:00&101 & -124& r$^3$&CIR\\
				49 &2013-06-07T03:00&31 & -78& r$^4$&CME\\
				50 &2014-02-28T00:00&239 & -97& r$^4$&CIR\\
                51 &2015-01-07T12:00&222 & -107& r$^1$&CME\\
                52 &2015-03-17T23:00&184 & -234$^\Lambda$& r$^3$&CME\\
                53 &2015-06-23T05:00&276 & -198$^\Lambda$& r$^4$&CME\\
                54 &2016-01-01T01:00&112 & -116& r$^3$&CME\\
                55 &2016-10-13T18:00&205 & -110& r$^3$&CME\\
                56 &2017-05-28T08:00&106 & -125& r$^2$&CME\\
                57 &2018-08-26T07:00& 211 & -175$^\Lambda$& r$^3$&CME\\
\botrule
\end{tabular}
\footnotetext{ r$^1$ - "very fast" recovery duration takes $\leq 4$ days.}
\footnotetext{ r$^2$ - "fast" recovery duration takes 4 to 6 days.}
\footnotetext{ r$^3$ - "slow" recovery duration takes 6 to 9 days.}
\footnotetext{ r$^4$ - "very slow" recovery duration takes $\geq 9$ days.}
\footnotetext[*]{ - Total recovery time from the (Dst)min to fully recovered.}
\footnotetext{$\Lambda$ - Intense GS group for Dst $\leq-150$ nT.}
\end{table}

\section{Results and Discussion}\label{sec:res}
Intense GSs are typically generated by CMEs, whereas CIRs only rarely generate strong storms.
 CMEs are responsible for 90\% of the events that were analyzed, while CIRs are responsible for 10\% of the events (see Table \ref{tabl1} last column) and two groups are considered from those driver based classification. \\
We categorized all of the selected GS events into four groups based on their recovery time, as described in Section \ref{sec:data}. The distribution of events across the groups is as follows: 28\% of events belong to the \textquotedblleft very-fast\textquotedblright recovery group (r$^1$), 24.5\% belong to the \textquotedblleft fast\textquotedblright recovery group (r$^2$), 23\% belong to the \textquotedblleft slow\textquotedblright recovery group (r$^3$), and 24.5\% belong to the \textquotedblleft very-slow\textquotedblright recovery group (r$^4$). Additionally, we classified the selected storms into four groups based on the time it takes for the storm main phase duration. The groups are as follows: \textquotedblleft very-fast\textquotedblright decrease events (d$_1$) with a decrease time of $ \leq 8$ hours, \textquotedblleft fast\textquotedblright decrease events (d$_2$) with a decrease time of 9 -- 12 hours, \textquotedblleft slow\textquotedblright decrease events (d$_3$) with a decrease time of 13 -- 22 hours, and \textquotedblleft very-slow\textquotedblright decrease events (d$_4$) with a decrease time of $\geq23$ hours \citep[][]{2024arXiv240203261A}. One unique group was also selected for the analysis, as described in Section \ref{sec:data} and footnote of Table \ref{tabl1}. As a result, we considered a total of 12 groups for this paper.\\

\begin{table}
\caption{List of $\tau$ from exponential fitting of recovery phase for geomagnetic index, IP SW parameters and their derivatives for various groupings as explained in the Sections \ref{sec:data} and \ref{sec:res}}.\label{t1}
\begin{tabular}{@{}lllllllll@{}}
\toprule
\textbf{Group}& \textbf{$\tau$(Dst)} & \textbf{R$^2$}&\textbf{$\tau$(Ey)} & \textbf{R$^2$}& \textbf{$\tau(\rho^{\frac{1}{3}}$v$^{\frac{4}{3}}$Bz)}& \textbf{R$^2$} & \textbf{$\tau$(B)} & \textbf{R$^2$ }\\
\midrule
$\mathrm{d_{5}}$& $34.48\pm0.68$ & 0.99 & $7.22\pm0$ & 0.83 & $7.51\pm0.47$ & 0.87  &$19.35\pm0.49$ &0.98\\
$\mathrm{d_{4}}$& $32.89\pm0.69$ & 0.99 & $12.68\pm0.83$ & 0.87 & $8.58\pm0.67$ & 0.81 &$18.91\pm0.61$ &0.96\\
$\mathrm{d}_{3}$& $30.56\pm0.57$ & 0.99 &$4.45\pm0.41$ & 0.76 & $5.12\pm0.41$ & 0.80 &$15.18\pm0.52$ &0.96\\
$\mathrm{d}_{2}$& $17.08\pm0.63$ & 0.95 &$4.00\pm0.80$ & 0.40 & $4.03\pm0.62$ & 0.53 &$21.43\pm0.77$ &0.96\\
$\mathrm{d}_{1}$& $20.15\pm0.4$ & 0.98 &$2.25\pm0.22$  & 0.74 & $2.04\pm0.18$ & 0.78 &$15.13\pm0.32$ &0.98\\
\hdashline[6pt/1.5pt]
$\mathrm{r^{4}}$& $43.22\pm0.54$ & 0.98 & $45.14\pm13.35$& 0.3 & $25.34\pm5.80$ & 0.38 &$21.85\pm0.92$ &0.94\\
$\mathrm{r}^{3}$& $29.21\pm1.13$ & 0.95 & $10.21\pm1.93$ & 0.43 & $11.20\pm1.58$ & 0.57 &$24.95\pm1.28$&0.92\\
$\mathrm{r}^{2}$& $34.73\pm1.13$ & 0.97 & $7.44\pm0.62$ & 0.79 &$13.44\pm1.13$ & 0.79 &$15.00\pm0.73$ &0.92\\
$\mathrm{r}^{1}$& $17.46\pm0.54$ & 0.96 &$3.47\pm0.50$ & 0.56 & $3.32\pm0.39$ & 0.66 & $20.15\pm0.73$&0.96\\
\hdashline[6pt/1.5pt]
            ICME& $34.77\pm0.73$ & 0.99 & $6.76\pm0.50$ & 0.83 &$6.81\pm0.42$ & 0.87 &$19.61\pm0.47$ &0.98\\
             CIR& $29.23\pm1.49$ & 0.93 &$3.12\pm0.57$ & 0.44 & $4.64\pm0.75$ & 0.50 & $17.42\pm1.33$&0.83\\
Int& 29.43$\pm$0.62 & 0.99 &3.16$\pm$0.00 & 0.80 & 3.84$\pm$0.26 & 0.85 & 16.38$\pm$0.58&0.96\\
\toprule  
\textbf{Group}  & $\tau$(\textbf{v}$\sigma$\textbf{B}) & \textbf{R$^2$}&$\tau(\sigma$\textbf{B}) & \textbf{R$^2$}& $\tau$(\textbf{Bz})& \textbf{R$^2$} & $\tau$(\textbf{v}) & \textbf{R$^2$ }\\
\midrule
$\mathrm{d}_{5}$  & $11.70\pm0.43$  & 0.95 &$12.71\pm0.42$  & 0.96  &  $5.95\pm0.51$& 0.78 &$1.3\mathrm{E}^{5}\pm1.3\mathrm{E}^{7}$ &0.96\\ 
$\mathrm{d}_{4}$  & $26.06\pm2.25$ & 0.81 &$27.96\pm2.69$  & 0.79 &  $13.15\pm0.98$& 0.83 &$1.1\mathrm{E}^{5}\pm2.2\mathrm{E}^{7}$ &0.77\\ 
$\mathrm{d}_{3}$  & $4.58\pm0.30$ & 0.86 &$8.22\pm0.68$  & 0.80 &  $3.72\pm0.27$& 0.83 &$164\pm47$ &0.85\\
$\mathrm{d_{2}}$  & $11.58\pm0.97$ & 0.79 & $8.88\pm0.77$ & 0.78 & $4.10\pm0.88$ & 0.36 &$5.6\mathrm{E}^{4}\pm3.3\mathrm{E}^{6}$ &0.93\\ 
$\mathrm{d_{1}}$  & $9.58\pm0.38$ & 0.94 & $9.00\pm0.41$ & 0.93 & $3.15\pm0.36$ & 0.67 &$265\pm50$ &0.97\\\hdashline[6pt/1.5pt] 
$\mathrm{r}^{4}$  & $43.78\pm6.05$ & 0.73 &$22.27\pm1.84$  & 0.82 &  $12.32\pm1.71$& 0.58 & $543\pm292$&0.94\\ 
$\mathrm{r}^{3}$  & $17.78\pm0$ & 0.84 & $8.93\pm0.55$ & 0.88 &$6.89\pm1.14$  & 0.49 &$1.8\mathrm{E}^{6}\pm-$ &0.89\\  
$\mathrm{r}^{2}$  & $12.50\pm0.91$ & 0.84 & $14.05\pm1.10$ & 0.82 & $8.58\pm0.64$ & 0.83 & $52.33\pm9.11$&0.68\\  
$\mathrm{r^{1}}$  & $5.04\pm0.24$ & 0.92 & $5.02\pm0.33$ &  0.86  & $4.05\pm0.68$ & 0.49 &$61.37\pm5.08$ &0.92\\  
\hdashline[6pt/1.5pt]
ICME& $10.34\pm0$ & 0.95 & $11.57\pm0.42$ & 0.95 &$8.27\pm0.63$ & 0.82 &$506\pm215$ &0.96\\
CIR& $19.10\pm1.33$ & 0.86 &$28.62\pm3.26$ & 0.73 & $3.15\pm0.54$& 0.47 & $148\pm9.27$&0.99\\
Int& 8.26$\pm$0.34 & 0.94 &10.86$\pm$0.43 & 0.95 & 5.59$\pm$0.47 & 0.79 & 71.51$\pm$4.28&0.97\\
\toprule  
\textbf{Group}  & \textbf{$\tau(\rho^{\frac{1}{3}}$v$\sigma$B)} & \textbf{R$^2$}&$\tau(\rho)$ & \textbf{R$^2$}& $\tau$(\textbf{P})& \textbf{R$^2$} & \textbf{$\tau(\rho^{\frac{1}{3}}$Ey)} & \textbf{R$^2$ }\\
\midrule
$\mathrm{d}_{5}$  &  $9.84\pm0.29$ & 0.97 &$10.88\pm0.65$  & 0.88  &  $11.50\pm0.26$& 0.98 &$7.90\pm0.57$ &0.83\\
$\mathrm{d}_{4}$  & $19.44\pm1.43$ & 0.84 &$14.29\pm1.85$  & 0.62 &  $15.14\pm0.98$& 0.87 &$8.64\pm0.64$ &0.82\\
$\mathrm{d}_{3}$  & $3.92\pm0.21$ & 0.90  &$12.16\pm0.61$  & 0.91 &  $9.48\pm0.39$& 0.94 &$6.11\pm0.45$ &0.83\\
$\mathrm{d_{2}}$  & $10.41\pm0.67$ & 0.87 & $7.74\pm0.67$ & 0.78 & $10.10\pm0.59$ & 0.89 &$4.09\pm0.64$ &0.51\\ 
$\mathrm{d_{1}}$  & $8.38\pm0.25$  & 0.97 & $8.85\pm0.64$ & 0.84 & $10.47\pm0.27$ & 0.98 &$2.20\pm0.20$ &0.76\\
\hdashline[6pt/1.5pt]  
$\mathrm{r}^{4}$  & $18.99\pm1.34$& 0.85 &$14.92\pm0.96$ & 0.87 &  $18.85\pm1.26$& 0.87 & $28.61\pm6.24$&0.42\\  
$\mathrm{r}^{3}$  & $9.56\pm0.44$ & 0.92 & $9.07\pm0.60$ & 0.86 &$9.70\pm0.24$  & 0.98 &$11.00\pm1.48$&0.60\\  
$\mathrm{r}^{2}$  & $14.45\pm0.87$& 0.88 & $11.91\pm1.56$ & 0.61 & $10.93\pm0.73$& 0.86 & $14.11\pm1.11$ &0.81\\  
$\mathrm{r^{1}}$  & $12.60\pm1.06$ & 0.79 & $3.59\pm0$ &  0.66  & $5.28\pm0.29$ & 0.90 &$4.05\pm0.47$ &0.66\\ 
\hdashline[6pt/1.5pt] 
           ICME   & $8.95\pm0.27$ & 0.97 & $10.00\pm0.61$ & 0.88 &$10.59\pm0.23$ & 0.98 &$7.43\pm0.48$ &0.86\\
            CIR   & $30.51\pm3.53$ & 0.73 &$15.90\pm1.26$ & 0.82 & $8.13\pm0.40$& 0.92 & $12.09\pm2.03$&0.49\\
Int & 9.22$\pm$0.31 & 0.96 &9.25$\pm$0.67 & 0.84 & 10.33$\pm$0.29 & 0.97 & 4.40$\pm$0.34&0.80\\
\toprule
\textbf{Group}  & $\tau$(\textbf{$\rho$v$\sigma$B}) & \textbf{R$^2$}&
$\tau$(\textbf{P$^{\frac{1}{2}}$Ey}) & \textbf{R$^2$}& \textbf{$\tau(\rho$Ey)}& \textbf{R$^2$} & $\tau$(\textbf{v$^{\frac{4}{3}}$Bz}) & \textbf{R$^2$}\\
\midrule
$\mathrm{d}_{5}$ & $8.07\pm0.24$  & 0.97 &$14.37\pm0.40$& 0.86 & $8.21\pm0.64$& 0.81 &$6.47\pm0.48$ &0.83\\ 
$\mathrm{d}_{4}$ & $18.22\pm1.67$ & 0.77 &$6.62\pm0.64$& 0.73 &  $5.07\pm0.49$& 0.74 &$12.51\pm0.83$ &0.86\\
$\mathrm{d}_{3}$ & $3.48\pm0.19$ & 0.90 &$4.05\pm0.22$  & 0.90 &  $8.63\pm0.76$& 0.77 &$3.83\pm0.39$ &0.72\\ 
$\mathrm{d_{2}}$ & $7.11\pm0.41$ & 0.88 & $3.92\pm0.54$ & 0.58 & $4.27\pm0.38$ & 0.76 &$3.97\pm0.78$ &0.41\\ 
$\mathrm{d_{1}}$ & $7.02\pm0.16$ & 0.98 & $1.63\pm0.13$ & 0.82 & $2.63\pm-$& 0.77 &$2.04\pm0.20$ &0.75\\ 
\hdashline[6pt/1.5pt] 
$\mathrm{r}^{4}$ & $12.25\pm0.79$ & 0.87 &$15.11\pm4.05$ & 0.23 & $0.07\pm504$& 0.63 &$42.87\pm13.37$&0.33\\ 
$\mathrm{r}^{3}$ & $6.62\pm0.19$ & 0.97 & $8.83\pm0.87$ & 0.73 &$11.33\pm1.15$  & 0.72 &$10.13\pm2.12$ &0.38\\ 
$\mathrm{r}^{2}$ & $10.72\pm0.62$ & 0.89 & $11.91\pm1.56$ & 0.61 & $10.20\pm0.94$ & 0.76 & $13.85\pm1.23$&0.78\\ 
$\mathrm{r^{1}}$ & $9.13\pm0.78$ & 0.78 & $3.45\pm0.27$ &  0.80  & $3.13\pm0.37$ & 0.66 &$3.30\pm0.43$ &0.61\\  
\hdashline[6pt/1.5pt]
ICME& $7.37\pm0.22$ & 0.97 & $5.40\pm0.36$ & 0.85 &$7.45\pm0.49$ & 0.86 &$5.95\pm0.43$ &0.83\\
 CIR& $20.52\pm2.70$ & 0.63 &$5.96\pm1.03$ & 0.46 & $5.38\pm0.79$& 0.55 & $3.11\pm0.60$&0.43\\
 Int& 7.36$\pm$0.25 & 0.96 &2.55$\pm$0.18 & 0.83 & 6.70$\pm$0.76 & 0.69 & 2.99$\pm$0.23&0.82\\
\botrule
\end{tabular}
\end{table}

\begin{table}

\begin{tabular}{@{}lllllllll@{}}
\toprule
\textbf{Group}  & $\tau$(\textbf{v$^{\frac{4}{3}}$B$ \rho^{\frac{1}{2}} $})  & \textbf{R$^2$}&$\tau$(\textbf{v$^{\frac{4}{3}}$Bz$\rho $}) & \textbf{R$^2$}& $\tau$(\textbf{v$^{\frac{4}{3}}$Bz$ \rho^{\frac{1}{2}} $})& \textbf{R$^2$} &$\tau$(\textbf{v$\sigma$B$\rho^{\frac{1}{2}}$})& \textbf{R$^2$} \\
\midrule
$\mathrm{d}_{5}$  & $15.58\pm0.23$  & 0.99 &$7.73\pm0.60$  & 0.81 & $7.67\pm0.49$& 0.86 &$9.30\pm0.26$ &0.97\\
$\mathrm{d}_{4}$  & $27.51\pm1.82$ & 0.89 &$0.06\pm1402$  & 0.62 &  $7.07\pm0.58$& 0.79 &$18.59\pm1.40$ &0.83\\
$\mathrm{d}_{3}$  & $12.18\pm0.40$ & 0.96 &$7.69\pm0.66$  & 0.78 &  $5.82\pm0.46$& 0.80 &$3.74\pm0.19$ &0.90\\
$\mathrm{d_{2}}$  & $14.93\pm0.43$ & 0.97 & $4.21\pm0.37$ & 0.77 & $4.04\pm0.02$ & 0.58 &$9.23\pm0.56$ &0.87\\
$\mathrm{d_{1}}$  & $11.19\pm0.18$ & 0.99 & $2.60\pm0.10$ & 0.79 & $2.02\pm0.18$ & 0.78 &$7.96\pm0.21$ &0.97\\
\hdashline[6pt/1.5pt]
$\mathrm{r}^{4}$  & $21.25\pm1.07$ & 0.92 &$0.07\pm388$  & 0.65 &  $21.71\pm4.82$& 0.38 &$16.05\pm1.01$ &0.87\\
$\mathrm{r}^{3}$  & $14.25\pm0.41$ & 0.97 & $11.53\pm1.20$ & 0.71 &$11.42\pm1.43$  & 0.63 &$8.41\pm0.32$ &0.95\\
$\mathrm{r}^{2}$  & $14.90\pm0.46$ & 0.97 & $8.24\pm0.82$ & 0.73 & $12.24\pm0.95$ & 0.81 &$13.13\pm0.76$ &0.89\\
$\mathrm{r^{1}}$  & $11.66\pm0.36$ & 0.97 & $3.3E^{-4}\pm0$ & 0.02& $3.72\pm0.37$ & 0.72 &$10.68\pm0.88$ &0.80\\
\hdashline[6pt/1.5pt]
              ICME& $7.37\pm0.22$ & 0.97 & $6.60\pm0.52$ & 0.81 &$6.89\pm0.43$ & 0.86 &$8.50\pm0.25$ &0.97\\
               CIR& $20.52\pm2.70$ & 0.63 &$7.71\pm0.82$ & 0.70 & $5.49\pm0.84$& 0.53 & $26.28\pm3.00$&0.71\\
Int& 12.07$\pm$0.24 & 0.99 &5.54$\pm$0.57 & 0.72 & 4.36$\pm$0.33&0.83 & 8.70$\pm$0.00&0.96\\
				\hline  
\textbf{Group}& \textbf{$\tau(\rho^{\frac{1}{2}}$Ey)} & \textbf{R$^2$}&\textbf{$\tau($P$^{\frac{1}{3}}$Ey)} & \textbf{R$^2$}& \textbf{$\tau($P$^{\frac{1}{2}}$v$^{\frac{4}{3}}$Bz)}& \textbf{R$^2$} & \textbf{$\tau$(P$^{\frac{1}{2}}$v$\sigma$B)} & \textbf{R$^2$ }\\
\midrule
$\mathrm{d_{5}}$& $8.24\pm0.54$ & 0.85 & $6.79\pm0.42$ & 0.87 & $3.45\pm0.34$ & 0.87 &$8.23\pm0.26$ &0.96\\
$\mathrm{d_{4}}$& $7.21\pm0.56$ & 0.81 & $8.64\pm0.64$ & 0.82 & $6.48\pm0.67$ &0.71 &$14.85\pm1.07$ &0.84\\
$\mathrm{d}_{3}$& $6.89\pm0.51$ & 0.82 &$4.23\pm0.26$ & 0.87 & $3.52\pm0.20$& 0.88 &$2.92\pm0.11$ &0.95\\
$\mathrm{d}_{2}$& $4.09\pm0.58$ & 0.57 &$3.98\pm0.61$ & 0.52 & $3.86\pm0.52$& 0.60 &$9.43\pm0.55$ &0.88\\
$\mathrm{d}_{1}$& $2.16\pm0.20$ & 0.76 &$1.79\pm0.15$  & 0.80 & $1.54\pm0.11$& 0.83 &$8.10\pm0.22$ &0.97\\
\hdashline[6pt/1.5pt]
$\mathrm{r^{4}}$& $25.14\pm5.32$ & 0.41 & $25.02\pm6.63$& 0.31 & $1.08\pm0.18$ & 0.53 &$19.26\pm1.32$ &0.86\\
$\mathrm{r}^{3}$& $11.18\pm1.34$ & 0.65 & $11.41\pm1.82$ & 0.51 & $12.05\pm2.00$ & 0.49 &$10.05\pm0.39$&0.94\\
$\mathrm{r}^{2}$& $13.32\pm0.97$ & 0.83 & $11.50\pm0.99$ & 0.78 &$7.25\pm0.78$ & 0.69 &$14.49\pm0.75$ &0.91\\
$\mathrm{r}^{1}$& $3.88\pm0.43$ & 0.68 &$3.17\pm4.3E^{-9}$ & 0.70 & $3.31\pm0.23$& 0.84 & $21.36\pm0.95$&0.82\\
\hdashline[6pt/1.5pt]
            ICME& $7.45\pm0.49$ & 0.85 & $6.03\pm0.38$ & 0.87 &$4.66\pm0.29$ & 0.87 &$7.71\pm0.25$ &0.96\\
            CIR& $5.38\pm0.79$ & 0.55 &$4.85\pm1.1E^{-8}$ & 0.46 & $6.06\pm1.09$& 0.44 & $9.80\pm0.56$&0.89\\
Int& 5.15$\pm$0.44 & 0.78 &2.81$\pm$0.00 & 0.83 & 2.50$\pm$0.17&0.87 & 5.35$\pm$0.20&0.95\\
				\hline  
\textbf{Group}& \textbf{$\tau($PEy)} & \textbf{R$^2$}&\textbf{$\tau($Pv$\sigma$B)}& \textbf{R$^2$}& \textbf{$\tau$(P$^{\frac{1}{3}}$v$^{\frac{4}{3}}$Bz)}& \textbf{R$^2$} & \textbf{$\tau$(P$^{\frac{1}{3}}$v$\sigma$B)} & \textbf{R$^2$ }\\
\midrule
$\mathrm{d_{5}}$& $3.48\pm0.30$ & 0.78 & $6.15\pm0.22$ & 0.95 & $6.01\pm0.36$ & 0.88 &$7.60\pm0.24$ &0.96\\
$\mathrm{d_{4}}$& $0.07\pm377$ & 0.65 & $9.16\pm0.76$ & 0.79 & $8.58\pm0.67$ &0.81 &$19.44\pm1.43$ &0.84\\
$\mathrm{d}_{3}$& $3.73\pm0.14$ & 0.95  &$2.16\pm0.05$ & 0.98 & $3.66\pm0.25$& 0.85 &$3.31\pm0.15$ &0.93\\
$\mathrm{d}_{2}$& $3.70\pm0.29$ & 0.81 &$6.58\pm0.37$ & 0.89& $3.93\pm0.59$& 0.54 &$10.78\pm0.68$ &0.87\\
$\mathrm{d}_{1}$& $1.27\pm0.08$ & 0.88 &$6.86\pm0.16$  & 0.98 & $1.67\pm0.14$& 0.81 &$8.58\pm0.26$ &0.97\\
\hdashline[6pt/1.5pt]
$\mathrm{r^{4}}$& $0.84\pm0.14$ & 0.58 & $14.58\pm1.10$& 0.83 & $19.99\pm5.52$ & 0.27 &$22.64\pm1.71$ &0.84\\
$\mathrm{r}^{3}$& $12.06\pm1.84$ & 0.54 & $7.51\pm0$ & 0.96 & $11.52\pm2.00$ & 0.47 &$11.34\pm0.53$&0.92\\
$\mathrm{r}^{2}$& $1.39\pm0$ & 0.76 & $10.76\pm0.45$ & 0.94 &$10.40\pm0.97$ & 0.75 &$16.13\pm0.93$ &0.89\\
$\mathrm{r}^{1}$& $2.35\pm0.12$ & 0.91 &$6.01\pm0.45$ & 0.82 & $3.02\pm0.29$& 0.74 & $16.04\pm1.32$&0.80\\
\hdashline[6pt/1.5pt]
            ICME& $2.61\pm0.22$ & 0.79 & $5.95\pm0.21$ & 0.96 &$5.28\pm0.32$ & 0.87 &$8.44\pm0.27$ &0.96\\
            CIR& $1.44\pm0.15$ & 0.73 &$4.69\pm0.27$ & 0.89 & $4.91\pm0.88$& 0.45 & $12.20\pm0.72$&0.88\\
Int& 1.85$\pm$0.14 & 0.82 &3.65$\pm$0.14 & 0.95 & 2.70$\pm$0.18&0.87 & 6.21$\pm$0.24&0.95\\
								\hline
\textbf{Group}& \textbf{$\tau($Pv$^{\frac{4}{3}}$Bz)} & \textbf{R$^2$}&& & & & & \\
\midrule
$\mathrm{d_{5}}$& $2.81\pm0.23$ & 0.80 & &  &  & &&\\
$\mathrm{d_{4}}$& $0.07\pm313$ & 0.65 &  &  & &&&\\
$\mathrm{d}_{3}$& $3.46\pm0.13$ & 0.95 & & & &  & &\\
$\mathrm{d}_{2}$& $3.61\pm0.27$ & 0.82 & &  & & &&\\
$\mathrm{d}_{1}$& $1.22\pm0.07$ & 0.89 & &  & & &&\\
\hdashline[6pt/1.5pt]
$\mathrm{r^{4}}$& $0.82\pm0.13$ & 0.63 & & & & & &\\
$\mathrm{r}^{3}$& $11.67\pm1.89$ & 0.50 &  & & &  &&\\
$\mathrm{r}^{2}$& $1.28\pm0.13$ & 0.76 &  &  & &  & &\\
$\mathrm{r}^{1}$& $2.26\pm0.09$ & 0.94 & & & & &&\\
\hdashline[6pt/1.5pt]
ICME& $2.32\pm0.18$ & 0.81 &  &  & & &&\\
CIR& $1.47\pm0.16$ & 0.72 & &  & & &&\\ 
Int& 1.87$\pm$0.13 & 0.85 &&  & & &&\\
\botrule
\end{tabular}
\footnotetext{d$_1$, \textquotedblleft very-fast\textquotedblright decline lasting less than 8 hours, d$_2$, \textquotedblleft fast\textquotedblright reduction lasting 9 to 12 hours.}
\footnotetext{d$_3$, \textquotedblleft slow\textquotedblright reduction that lasts between 13 and 22 hours.}
\footnotetext{d$_4$, \textquotedblleft very-slow\textquotedblright decline lasting more than 23 hours, d$_5$, The entire data set.}
\footnotetext{Int, \textquotedblleft intense\textquotedblright storms Dst $\leq-150$ nT.}

\end{table}
\subsection{Fitting dynamics of characteristics time}\label{curve}
\begin{figure}
   	\centering
  	\includegraphics[width=\textwidth]{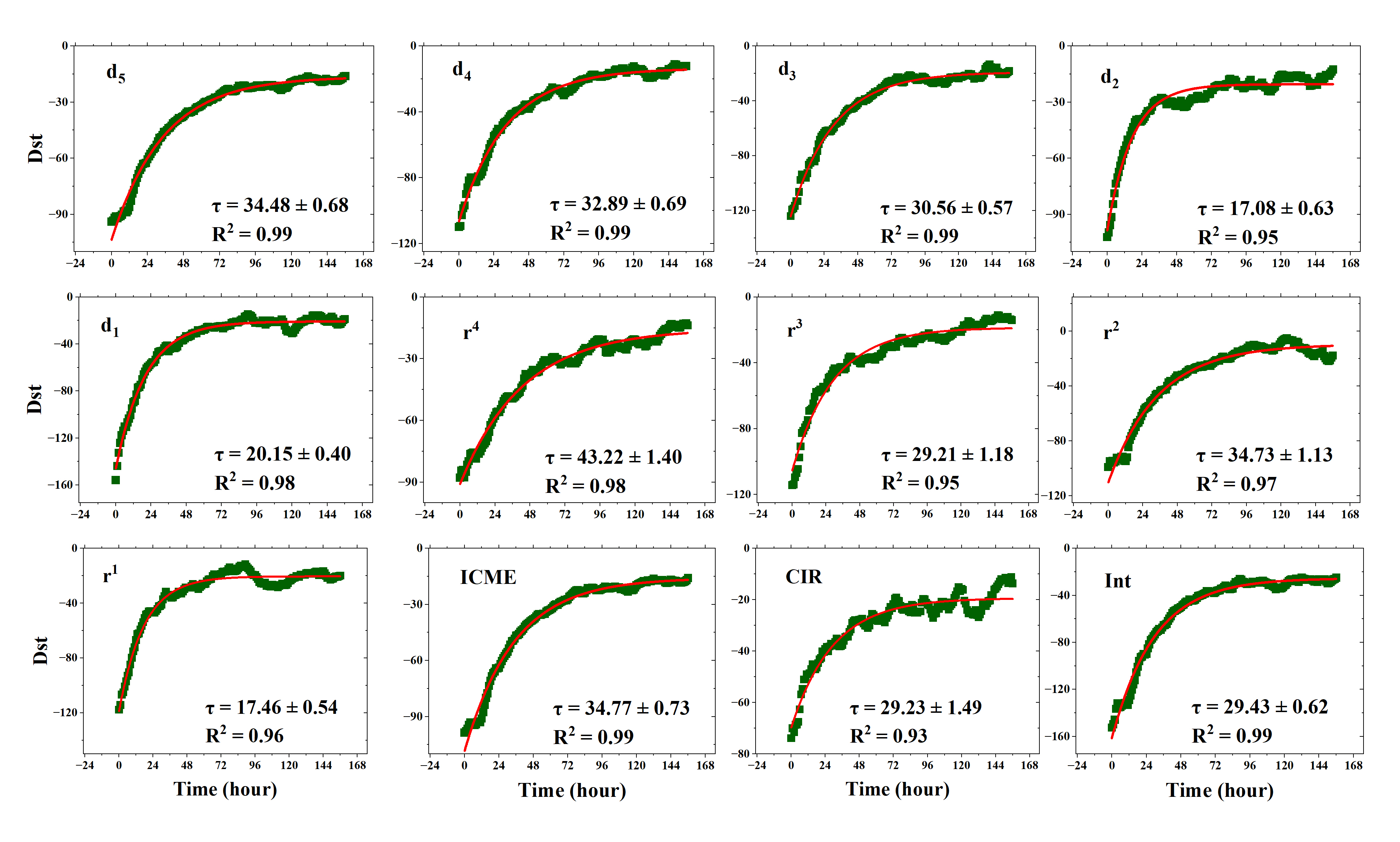}
\caption{shows the scatter plot with exponential fit ($y = 
  y_0+Ae^{-(x-x_0)/\tau}$) to the recovery phase of superposed epoch result of Dst (nT); all 57 GSs d$_5$(top-left), \textquotedblleft very-slow\textquotedblright dip group d$_4$ (top second column), \textquotedblleft slow\textquotedblright dip group d$_3$ (top third column),  \textquotedblleft fast\textquotedblright dip group d$_2$ (top-right), \textquotedblleft very-fast\textquotedblright dip group d$_1$ (second row right), groups r$^4$ (second row second column), r$^3$ (second row third column), r$^2$ (second row right), r$^1$ (bottom-first column), CME (bottom second column), CIR (bottom third column) and Int (bottom right). The beginning of the storm recovery is represented by zero hour in superposed charts. The red colored line shows the model curve that best depicts how our data should be fitted, while the green colored boxes represents the Dst values.}\label{fig:Dst}  
   \end{figure}

\begin{figure}
   	\centering
  	\includegraphics[width=\textwidth]{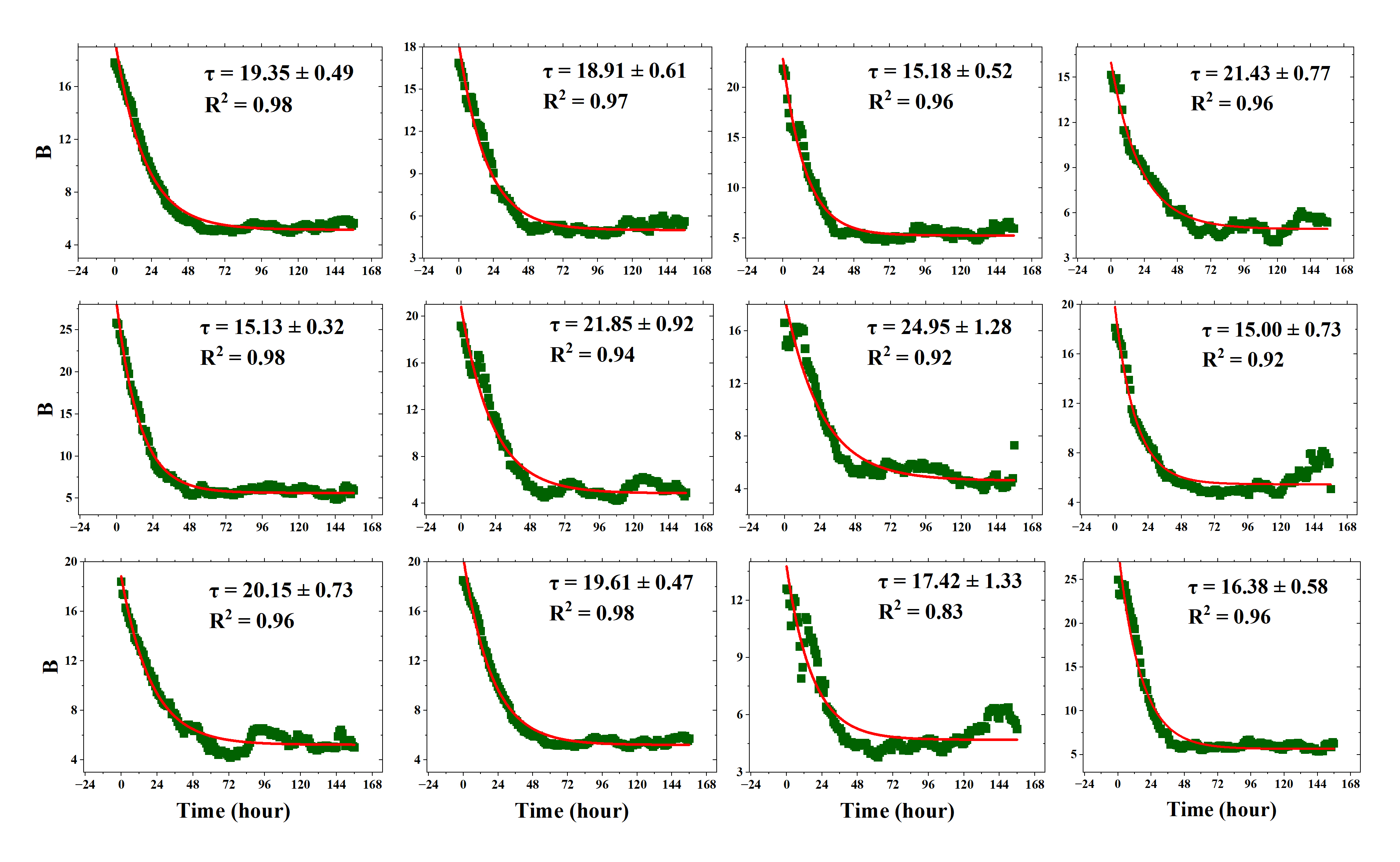}
\caption{Exponential fit to IMF (B) decay profile during recovery phase on respective groups adopted from Figure \ref{fig:Dst}}\label{fig:B}
  
   \end{figure}

\begin{figure}
   	\centering
  	\includegraphics[width=\textwidth]{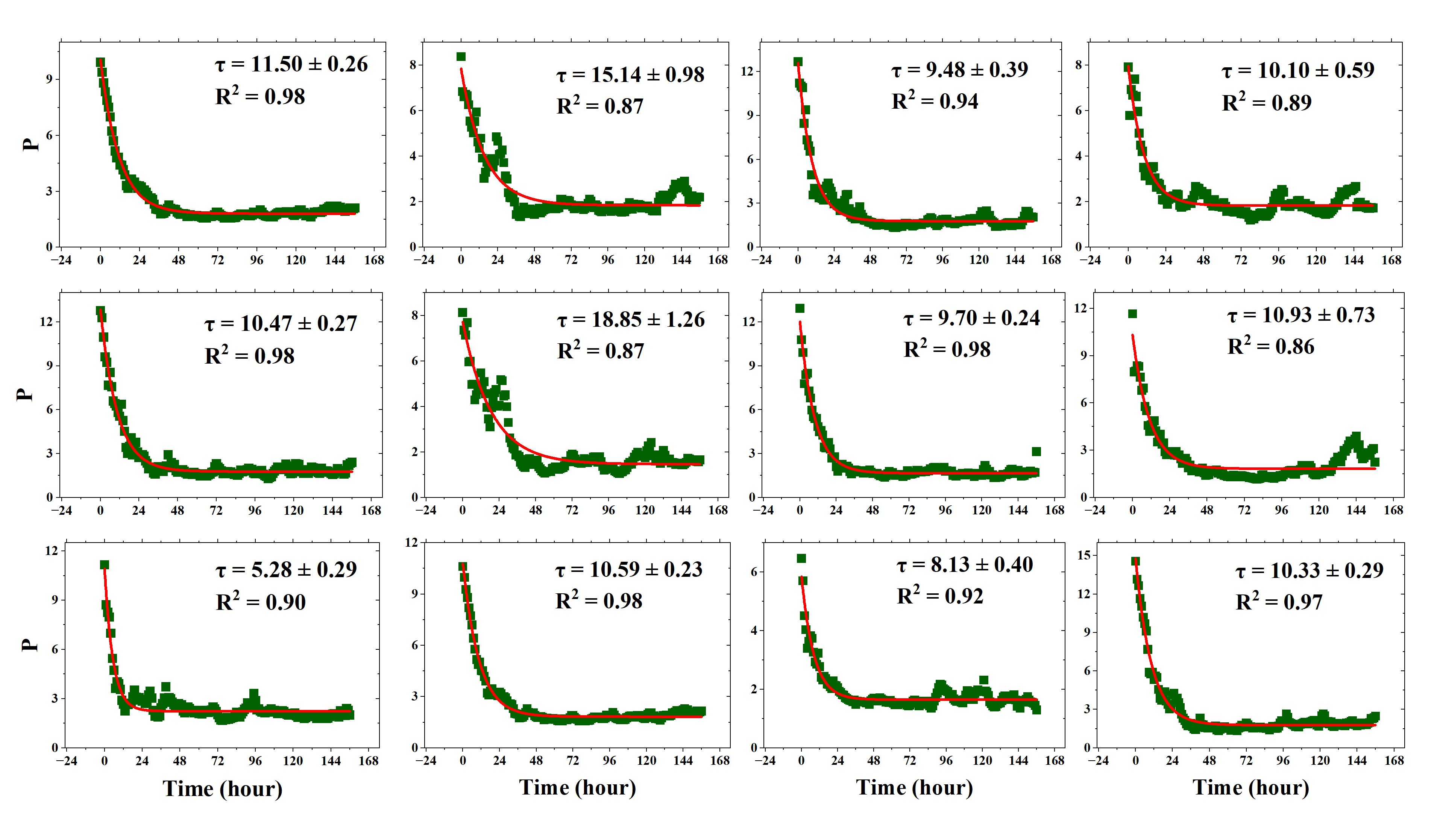}
\caption{Exponential fit to dynamic pressure (P) decay profile during recovery phase on respective groups adopted from Figure \ref{fig:Dst}}\label{fig:P}
  
   \end{figure}

\begin{figure}
   	\centering
  	\includegraphics[width=\textwidth]{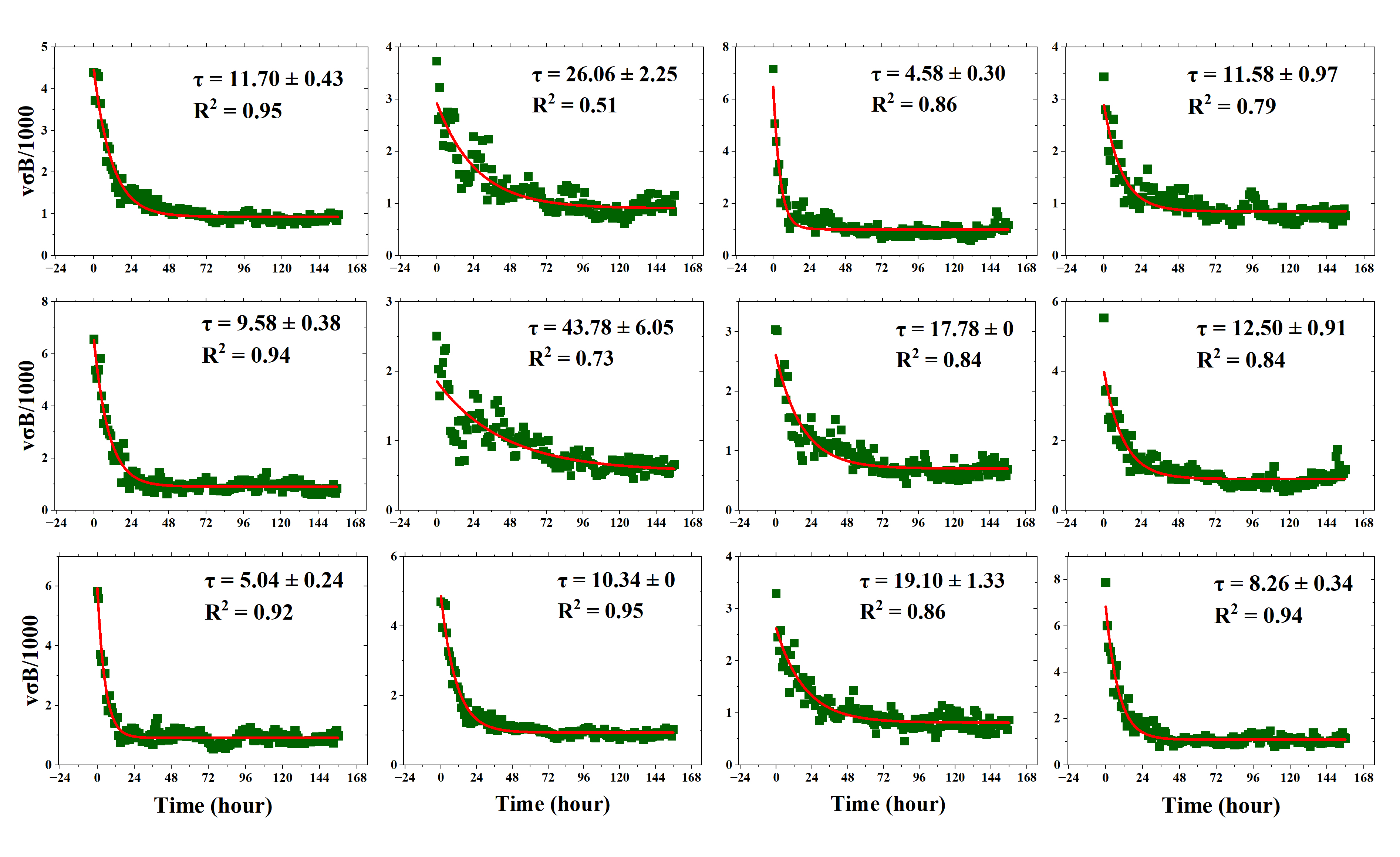}
\caption{Exponential fit to CF (v$\sigma$B) decay profile during recovery phase on respective groups adopted from Figure \ref{fig:Dst}}\label{fig:vsBv}
  
   \end{figure}

\begin{figure}
   	\centering
  	\includegraphics[width=\textwidth]{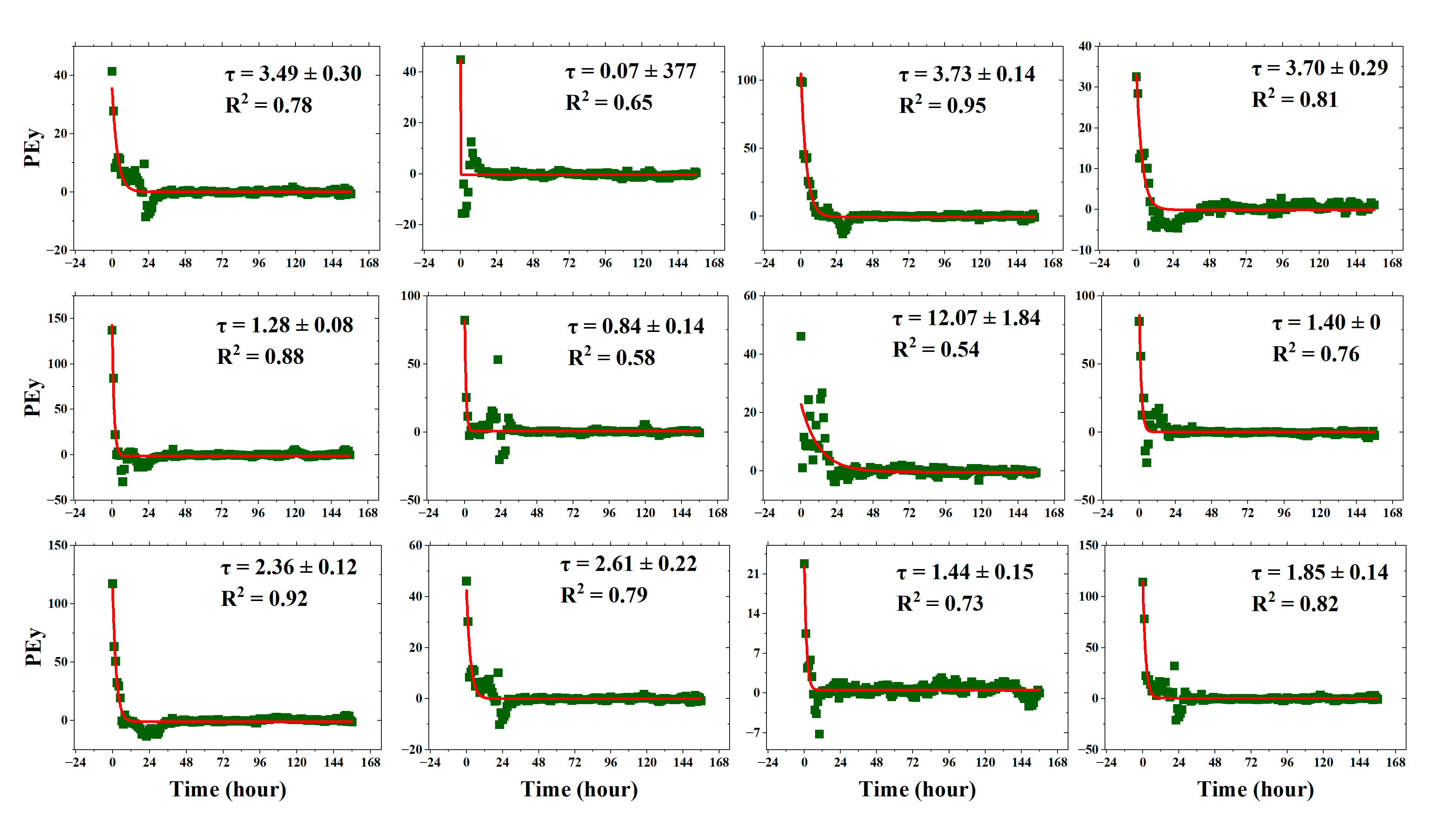}
\caption{Exponential fit to CF (PEy) decay profile during recovery phase on respective groups adopted from Figure \ref{fig:Dst}}
\label{fig:PEy}
  
   \end{figure}

\begin{figure}
   	\centering
  	\includegraphics[width=\textwidth]{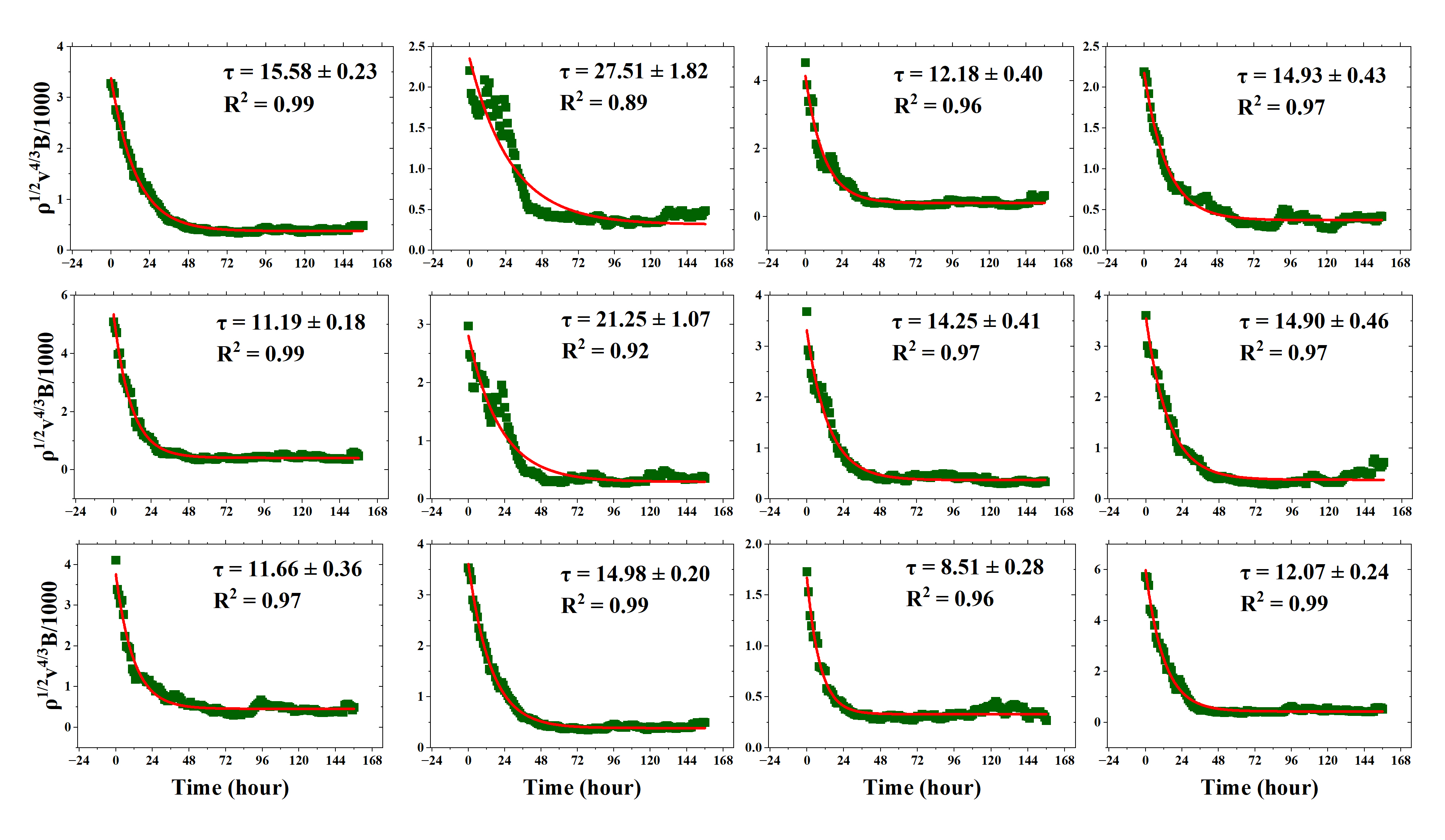}
\caption{Exponential fit to CF (P$^{\frac{1}{3}}$v$\sigma$B) decay profile during recovery phase on respective groups adopted from Figure \ref{fig:Dst}}
\label{fig:p13vsBv}
  
   \end{figure}

\begin{figure}
   	\centering
  	\includegraphics[width=\textwidth]{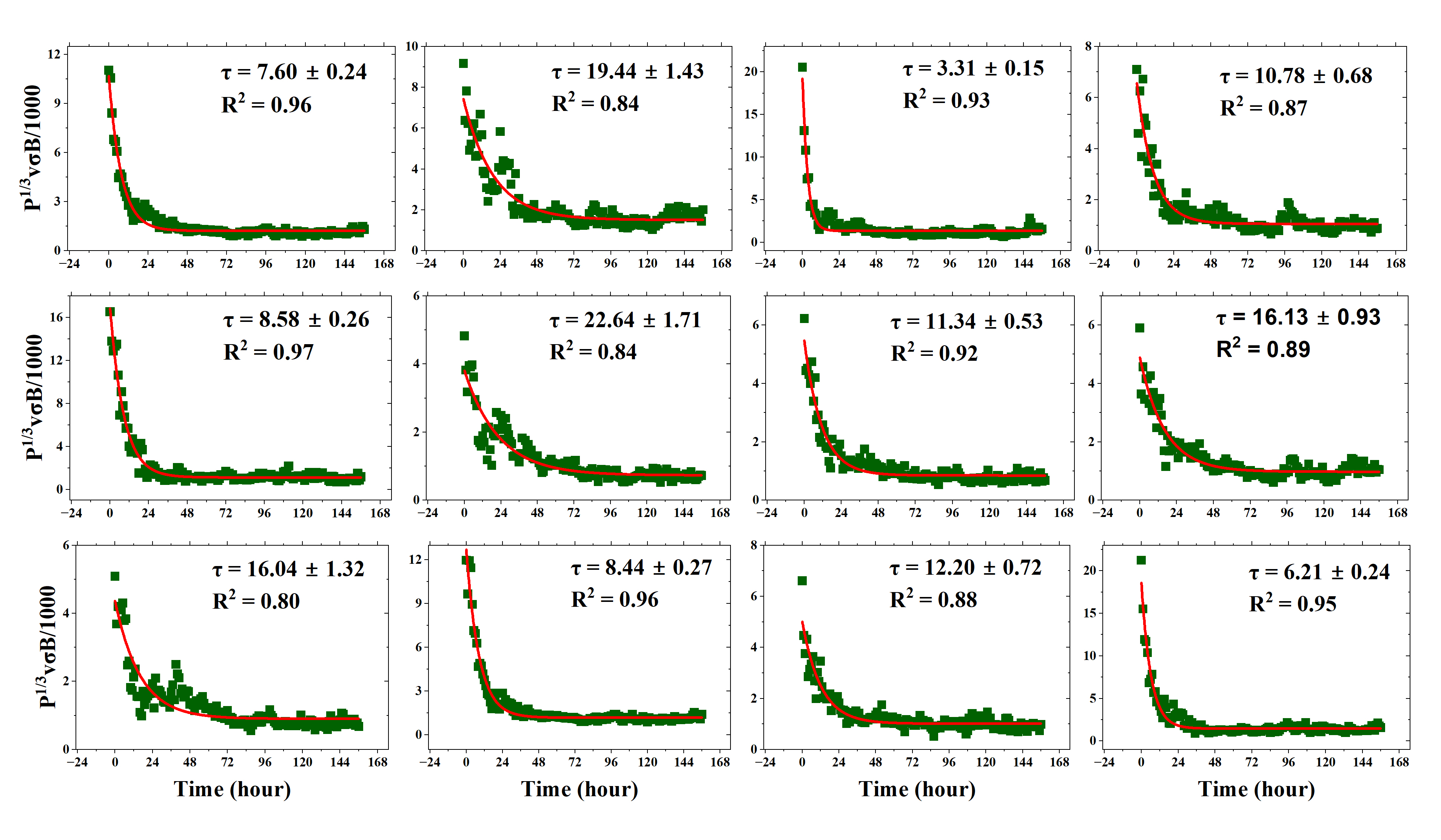}
\caption{Exponential fit to CF ($\rho^{\frac{1}{2}}$v$^{\frac{4}{3}}$B) decay profile during recovery phase on respective groups adopted from Figure \ref{fig:Dst}}\label{fig:n12v43B}
   \end{figure}

We employed the superposed epoch approach to analyze Dst, IP SW parameters, and their functions, which were divided into three categories:
\begin{itemize}
	\item A - a single IP SW parameter that includes (B, Bz, P, $\rho$, v, $\sigma$B \& Ey)
	\item B - Function of two IP parameters, of viscous terms ($\rho$, P, v terms) and electric field related terms (v$\sigma$B, PEy, P$^{\frac{1}{2}}$Ey, P$^{\frac{1}{3}}$Ey, $\rho^{\frac{1}{2}}$Ey, $\rho^{\frac{1}{3}}$Ey, $\rho$Ey, \& v$^{\frac{4}{3}}$Bz)
	\item C - Functions of three IP SW parameters which comprises viscus ($\rho$, P, v terms) and electric field related terms (v$^{\frac{4}{3}}$Bz$\rho$, v$^{\frac{4}{3}}$Bz$\rho^{\frac{1}{2}}$, v$^{\frac{4}{3}}$Bz$\rho^{\frac{1}{3}}$, v$^{\frac{4}{3}}$BzP$^{\frac{1}{3}}$, v$^{\frac{4}{3}}$BzP$^{\frac{1}{2}}$, v$^{\frac{4}{3}}$BzP, Pv$\sigma$B, P$^{\frac{1}{2}}$v$\sigma$B, P$^{\frac{1}{3}}$v$\sigma$B, $\rho^{\frac{1}{2}}$v$\sigma$B, $\rho^{\frac{1}{3}}$v$\sigma$B, $\rho$v$\sigma$B, \& v$^{\frac{4}{3}}$B$\rho^{\frac{1}{2}}$)
\end{itemize}
In addition to Dst, we have performed exponential curve fit to all parameters and functions (mentioned in class A, B \& C above) and the results were tabulated in Table \ref{t1}. However, for graphical presentation we have taken two representative parameters or functions from each class. To that end, IMF B and dynamic pressure P have been considered from class$-$A of single parameters (see, Figures \ref{fig:B} \& \ref{fig:P}), while v$\sigma$B and PEy have been taken from class$-$B of two parameter derived CFs (see Figures \ref{fig:vsBv} \& \ref{fig:PEy}). Three parameter derived CFs taken from class$-$C are v$^{\frac{4}{3}}$B$\rho^{\frac{1}{2}}$ \& P$^{\frac{1}{3}}$v$\sigma$B which have been shown in (Figures \ref{fig:p13vsBv} \& \ref{fig:n12v43B}).\\
The exponential curve fit plots for the recovery profiles of each of the geomagnetic index, IP SW parameters/functions, as shown in Figures \ref{fig:Dst} through \ref{fig:n12v43B}, are the results of superposed analysis for each group discussed in previous Sections.
Figure \ref{fig:Dst} is the geomagnetic index Dst, and Figures (\ref{fig:B} -- \ref{fig:n12v43B}) are the selected best two parameters from each class (A, B \& C) represent the recovery profile with each figure has 12 panels for 12 different groups namely: the entire set of all 57 GSs d$_5$(top-left), \textquotedblleft very-slow\textquotedblright dip group d$_4$ (top second column), \textquotedblleft slow\textquotedblright dip group d$_3$ (top third column),  \textquotedblleft fast\textquotedblright dip group d$_2$ (top-right), \textquotedblleft very-fast\textquotedblright dip group d$_1$ (second row left), groups r$^4$ (second row second column), r$^3$ (second row third column), r$^2$ (second row right), r$^1$ (bottom-left), ICME (bottom second column), CIR (bottom third column) and Int (bottom-right).\\
The Dst recovery profile can be approximated by exponential time constant ($\tau$), which is dependent on the minimum Dst index and varies with time \citep[e.g.,][]{2008AnGeo..26.2543M, 2012JGRA..117.8207Y}. We adopt exponential curve fit to describe the the ring current decay for each group.
Those Figures depict time evolution plots of storm recovery profiles and fitted with an exponential curve for all the selected parameters/functions. The curve fit equation is: $\mathrm{y = y_0+Ae^{-(x-x_0)/\tau}}$. 
y$_0$ is the GS at its peak which is taken as zero hour for recovery phase.
Our focus is on the time constant $\tau$, which provides information on the rate of the storm recovery and shown inside each of the figures as well as tabulated in Table \ref{t1}. 
The scattered green boxes are our data points and the red line shows the best fit.\\ 
It has been reported \citep[e.g., see,][and references therein]{2014JGRA..119.8126Y, 2016Ap&SS.361..253B, 2019E&SS....6.2000P} that CMEs are generally identified as the primary origins of significant GSs, whereas CIRs are more frequently attributed to moderate and mild magnetic storms and that the recovery profile of ICME is smooth while that of CIR recovery fluctuates which is consistent with our result shown in Figure \ref{fig:Dst}.
\cite{Bothmer1995TheIA, 2006JGRA..111.7S08B, 2011Ap&SS.331...91M} reported that strong GSs with persistent IMF Bz variations are facilitated by CMEs. On the other hand, CIRs are related to milder and more moderate storms and exhibit significant Bz fluctuations. This observation is in line with the results of our finding.
\\ 
The recovery phase of GS has two steps: faster first step and slower second step \citep[e.g.,][]{1990P&SS...38.1133T, 1997GMS....98..107D, 2011JASTP..73.1831M}.
From the time characteristics of group d$_{2}$, (Dst)$_\mathrm{min}=-102$ nT, it is evident that it recovers within 17 hours. It indicates the correlation between storm intensity and recovery duration. Our result shows slower recovery duration than previous studies \citep[e.g.,][]{1975JGR....80.4204B} and \citep[][]{2002JGRA..107.1059D}, who reported 7.7 and $14\pm4$ hours, respectively. Therefore, more intense storms relatively takes longer duration to return its onset level. In addition to intensity of storms, the main phase duration has direct relationship with the recovery time. The longer storm main phase takes longer duration to recover from disturbance. 
Storms with short main phase durations shows in general two stages of recovery: a rapid first step, which takes about 30 hours for groups of d$_{1}$, d$_{2}$ and r$^{1}$, followed by a gradual second step recovery.
From this results we can deduce that the ring current decays faster for storms with short main phase durations in general and it decays quickly for one day followed by slow long-lasting second-step recovery. On the other hand, storms with longer main phase duration show one step very slow steady recovery.
\\
In the same manner as Figure \ref{fig:Dst}, Figure \ref{fig:B} 
shows exponential decay of single parameter IMF magnitude B recovery profile with the quickest and smoothest recovery group r$^2$ and a time constant of 15 hours. In contrast, r$^4$ exhibits a slow and uneven distribution with a characteristic time of 22 hours (see, Figure \ref{fig:B}).
The decay recovery dynamics on average take one day steep/sharp recovery which takes shorter duration than Dst.
Figure \ref{fig:P} shows dynamic pressure (P) which follows similar decay dynamics with IMF B, takes very short recovery duration $\simeq$ 12 hours for r$ ^{1} $ and around one day for other groups. P has been fully recovered during 32\% of the first step ring current decay. If the peak of P coincides with Dst minimum, it returns to quiet level with in the initial step of ring current decay, implies that the effect of P on the recovery of storms is only for few hours. Notably, the P effect on the ring current decay of magnetic storms are neglected.\\
Figure \ref{fig:vsBv} is a dual parameter CF (v$\sigma$B/1000) which shows similar recovery profile with P. It takes (5 -- 44) hours to recover for fast and slow groups as (d$_3$ \& r$^4$) respectively.
Figure \ref{fig:PEy} is the dual parameter CF (PEy) which shows similar dynamics with v$\sigma$B/1000.
Figure \ref{fig:p13vsBv} is triple parameter CF (P$^{\frac{1}{3}}$v$\sigma$B/1000) which shows similar recovery morphology with v$\sigma$B/1000.
In general, Figures (\ref{fig:vsBv} \& \ref{fig:p13vsBv}) shows that the turbulent Sigma IMF vector ($\sigma$B) combined with IP SW speed (v) and P which signifies the level of fluctuations/turbulence during recovery phase of those structures. Especially, Figure \ref{fig:vsBv} which has a 1:1 combination of Sigma IMF vector ($\sigma$B) and v gives valuable information about the turbulence behavior storms during recovery period. Explicitly, intense GSs fluctuate for short time and less intense storms continue their recovery with turbulence for longer periods. Storms triggered by CME (where its intensity is almost double to CIR) is recovered during 50\% of CIR recovery period. From this result we can suggest that during recovery of very intense storms as group 'Int' (Dst $\leq$ -150 nT) only the first initial phase (28\%) of recovery period has perturbation as seen from Figures \ref{fig:Dst} and \ref{fig:vsBv}. Intense magnetic storms are driven by CMEs with the long-duration south-ward IMF component Bz \citep[][]{1999SSRv...88..529G, 2008JGRA..113.5221E}.
From our finding we observed that, the recovery characteristics of CF v$\sigma$B, for CME and CIR groups together with the same groups of Dst, CIR storms remain with fluctuations for the long period of time, which is supported by earlier studies \citep[e.g.,][]{2006ilws.conf..266G, 2011Ap&SS.331...91M}.\\
Figure \ref{fig:n12v43B} a triple parameter CF (v$^{\frac{4}{3}}$B$\rho^{\frac{1}{2}}/1000$) is a combination of viscus term $\rho$ and electric field related function v$^{\frac{4}{3}}$B, shows exponential decay recovery dynamics which returns to quiet level earlier than Dst.
From all IP SW parameters/functions, single parameters (Figures \ref{fig:B} \& \ref{fig:P}), dual parameter CFs (Figures \ref{fig:vsBv} \& \ref{fig:PEy}) and triple parameter CFs (Figures \ref{fig:p13vsBv} \& \ref{fig:n12v43B}) the CME driven storms recovery duration are shorter than the CIRs in (B, P and v$^{\frac{4}{3}}$B$\rho^{\frac{1}{2}}/1000$), where as for the rest CIR triggered storms recovered fast.\\
In addition to the displayed seven figures, we have analyzed the rest 22 IP SW parameters/functions, similarly with the same exponential function and their recovery characteristic follows either decay/growth and all implying values are tabulated in Table \ref{t1}. Here we have present parameters which represent best recovery fits from each class (A, B \& C) as (single, dual \& triple functions) classifications respectively.
We used derived CFs from the previous works \citep[e.g.,][]{2007JGRA..112.1206N, 2024arXiv240203261A} and used other functions to test their correlations with the recovery characteristics and morphology of GSs. In this work we used combinations of viscus term ($\rho$) and dynamic pressure term (P) with different power indices (1, $\frac{1}{2}$ \& $\frac{1}{3}$) combined with electric field related functions (IP dawn-dusk electric field -vBz and terrestrial electric field vB) which are also functions of IP SW speed v with different powers (1, $\frac{4}{3}$) and IMF (magnitude B, north-south component Bz \& Sigma vector $\sigma$B which shows turbulence). Such CFs are important to get information not only in the main phase of GSs, but also for the recovery phases of GSs. Our finding confirms that, the derived electrodynamical SW-magnetosphere CFs of viscus term ($\rho$) with electric field related functions are very important indicators of recovery phase of GSs.
As we can see from Table \ref{t1} derived functions PEy and Pv$^{\frac{4}{3}}$Bz are the two fastest recovered parameters/functions whereas IP SW speed v and IMF B show slow recovery. Even velocity didn't recover fully with in our considered time intervals.

%
%
%
 \subsection{Relation between Dst and IP SW parameters recovery time constants}\label{sec:dynamics}
\begin{figure}
    \includegraphics[width=\textwidth]{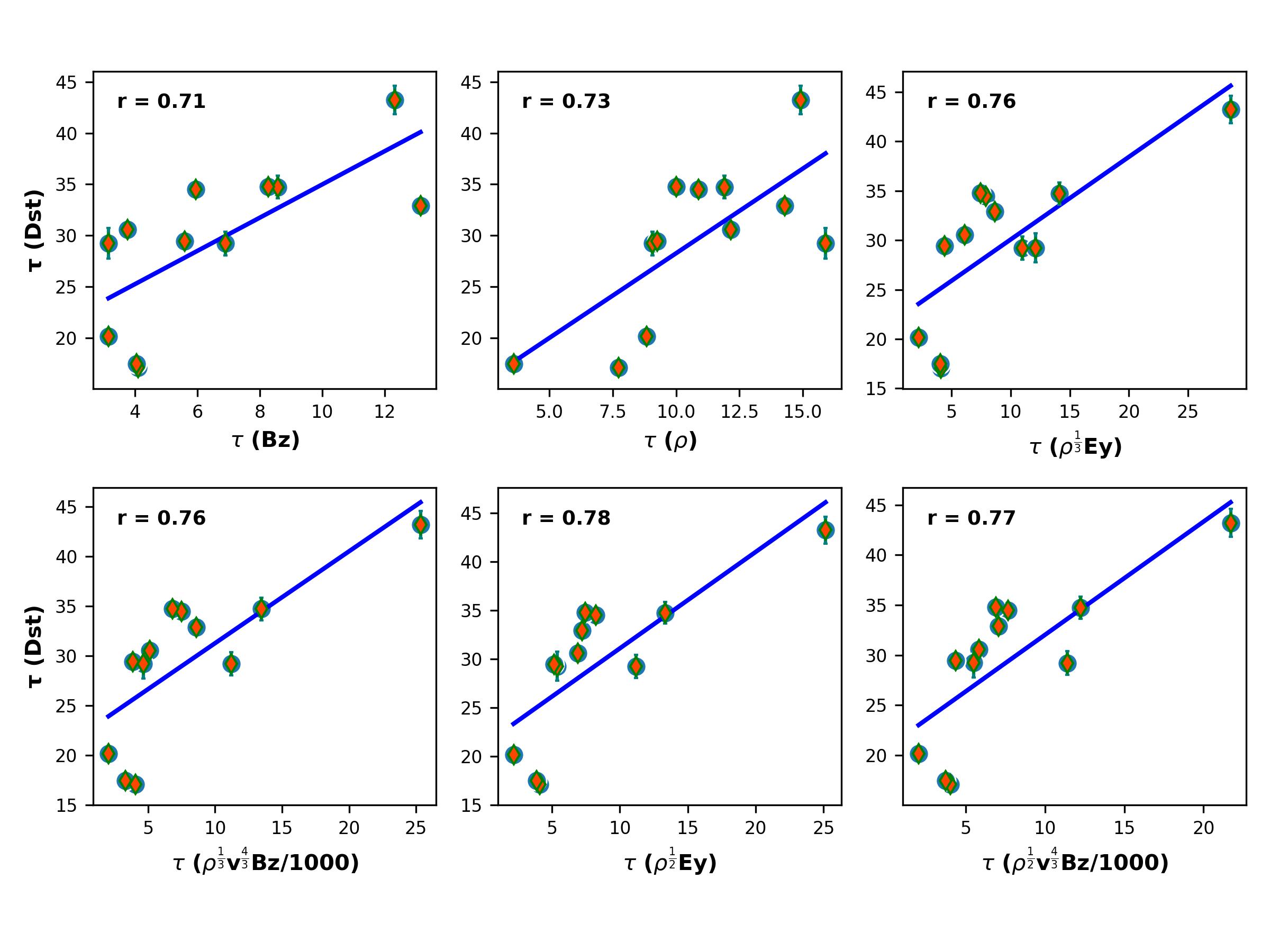}
   \caption{The scatter error bar plots depict the relation between recovery time constant ($\tau$) of Dst and representative IP SW parameters and their derivatives}. 
\label{fig:dstcorr}
\end{figure}

The impact of different SW parameters on the generation of GSs has been investigated in many previous studies, \cite[e.g.,][]{Richardson2007CorrectionT, Tsurutani1992GreatMS}.
Figure (\ref{fig:dstcorr}) show scatter error bar plots and linear regression analyses depicting the relationships between Dst and numerous IP SW parameters and product CFs. These visualizations were generated based on the recovery time constant ($\tau$) derived from the superposed epoch analysis (see Table \ref{t1}). The data points are represented by red boxes, accompanied by green bars denoting errors, and the blue lines indicate the linear fit model. \\
Figure \ref{fig:dstcorr} depicts the best-two IP SW plasma and field parameters/functions for each class (single, double, and triple) as explained in Section \ref{curve}.
Six panels are shown (the north-south component of IMF Bz and IP proton density $\rho$) 
selected as best-two correlated single parameters, ($\rho^{\frac{1}{2}}$Ey \& $\rho^{\frac{1}{3}}$Ey) as best-two dual parameter CFs and (v$^{\frac{4}{3}}$Bz$\rho^{\frac{1}{2}}$/1000 \& v$^{\frac{4}{3}}$Bz$\rho^{\frac{1}{3}}$/1000) are taken to be best-two triple parameter CFs. 
With some limitation of spread in scatter plots, we can conclude that the double and triple derived CFs, which are combinations of the square root of the viscus term ($\rho^{\frac{1}{2}}$) and the electric field-related functions (Ey \& v$^{\frac{4}{3}}$Bz) of ($\rho^{\frac{1}{2}}$Ey) and v$^{\frac{4}{3}}$Bz$\rho^{\frac{1}{2}}$/1000), respectively, are highly correlated with the recovery time constant of Dst. Notably, $\rho^{\frac{1}{2}}$Ey is the best function to determine the recovery time characteristics of GSs as shown in Table \ref{tt}.\\
With a Pearson's correlation coefficient of r$=0.78$, the time constant of the dual parameter CF ($\rho^{\frac{1}{2}}$Ey exhibits the highest correlation with Dst. The best-fit equation is $\tau(\mathrm{Dst})=(0.99\pm0.84)$($\rho^{\frac{1}{2}}$Ey)+($21.16\pm1.01$).
Overall, the results revealed significant correlations, indicating the interdependence and influence of various IP parameters/functions on the recovery time profiles. Distinctively, combination functions of viscus term ($\rho$) and electric field functions such as ($\rho^{\frac{1}{2}}$Ey and v$^{\frac{4}{3}}$Bz$\rho^{\frac{1}{2}}$) played crucial roles in determining the recovery time constant of Dst, which is an improvement to previous works by \cite{OBrien2000AnEP} who have reported that, the ring current decay is best described by electric field related function vBs (-vBz). The rest of all single, dual and triple parameter functions have underwent linear correlation analysis and the results are tabulated in Table \ref{tt}.

\begin{table}
\caption{Presents the Pearson's linear correlation coefficient and slope obtained from linear regression analyses conducted between the recovery time constants ($\tau$) for Dst and IP SW parameters, along with their product functions. Figure \ref{fig:dstcorr} provide visual representations of these linear relationships for selected parameters. }\label{tt}
\begin{tabular*}{\textwidth}{@{\extracolsep\fill}lccccccc}
\toprule%
\textbf{No}& \textbf{Param.}&\multicolumn{2}{@{}c@{}}{\textbf{Dst (nT)}} &\textbf{No}&\textbf{Param.}& \multicolumn{2}{@{}c@{}}{\textbf{Dst (nT)}} \\\cmidrule{3-4}\cmidrule{7-8}%
& & \textbf{r} & \textbf{s} & & & \textbf{r} & \textbf{s} \\
\midrule
1&B &0.04  & 0.09$\pm$3.06&15&P$^{\frac{1}{2}}$v$^{\frac{4}{3}}$Bz/1000 &0.10  &0.27$\pm$3.22 \\
2&Bz &0.71  & 1.62$\pm$1.88&16&P$^{\frac{1}{3}}$v$^{\frac{4}{3}}$Bz/1000 &0.72  & 1.08$\pm$1.45\\
3&$\sigma$B &0.52  & 0.51$\pm$0.97&17&$\rho^{\frac{1}{2}}$v$^{\frac{4}{3}}$B/1000 &0.48  &0.73$\pm$1.28 \\
4&$\rho$ &0.73  &1.65$\pm$1.23&18&$\rho^{\frac{1}{3}}$v$^{\frac{4}{3}}$Bz/1000 &0.76 &0.92$\pm$1.16 \\
5&v &0.01  &0.01$\pm$0.001&19&$\rho^{\frac{1}{2}}$v$^{\frac{4}{3}}$Bz/1000 &0.77 &1.13$\pm$1.32 \\ 
6&P &0.71  &1.64$\pm$1.24&20&$\rho$v$^{\frac{4}{3}}$Bz/1000 &0.15  &0.30$\pm$2.52\\
7&v$\sigma$B/1000 &0.62 & 0.44$\pm$0.71& 21&Ey &0.66 &0.44$\pm$0.73 \\
8&$\rho$v$\sigma$B/1000 &0.24 &0.38$\pm$1.77& 22&PEy &-0.13  &-0.32$\pm$3.32 \\ 
9&$\rho^{\frac{1}{2}}$v$\sigma$B/1000 &0.25 & 0.32$\pm$1.46&23&$\rho$Ey &0.16  &0.38$\pm$2.82  \\
10&$\rho^{\frac{1}{3}}$v$\sigma$B/1000 &0.23 &0.26$\pm$1.27&24&$\rho^{\frac{1}{2}}$Ey &0.78 &0.99$\pm$1.17 \\ 
11&Pv$\sigma$B/1000 &0.48 &1.13$\pm$2.25&25&$\rho^{\frac{1}{3}}$Ey &0.76 &0.84$\pm$1.04 \\
12&P$^{\frac{1}{2}}$v$\sigma$B/1000 &0.35 &0.62$\pm$1.79&26&P$^{\frac{1}{2}}$Ey &0.70 & 1.19$\pm$1.57 \\
13&P$^{\frac{1}{3}}$v$\sigma$B/1000 &0.29 &0.40$\pm$1.49&27&P$^{\frac{1}{3}}$Ey &0.73 & 0.90$\pm$1.21\\
14&Pv$^{\frac{4}{3}}$Bz/1000 &-0.14 &-0.37$\pm$3.45&28&v$^{\frac{4}{3}}$Bz/1000 &-0.14 &-0.37$\pm$3.45 \\
\botrule
\end{tabular*}
\footnotetext[]{\textbf{r}, represents Pearson's correlation coefficient is a measure of linear relation between}
\footnotetext[]{geomagnetic index and other parameters ($r= \frac{\Sigma(x_i-\overline{x})(y_i-\overline{y})}{\sqrt{\Sigma(x_i-\overline{x})^2\Sigma(y_i-\overline{y})^2}})$}
\footnotetext[]{\textbf{s}, stands for the slope of the linear regression as $y=sx+i$.}
\end{table}

 \subsection{Dynamical relationship between Dst and IP SW parameters during recovery phase}
\label{sec:index vs ip}
\begin{figure}
    \includegraphics[width=\textwidth]{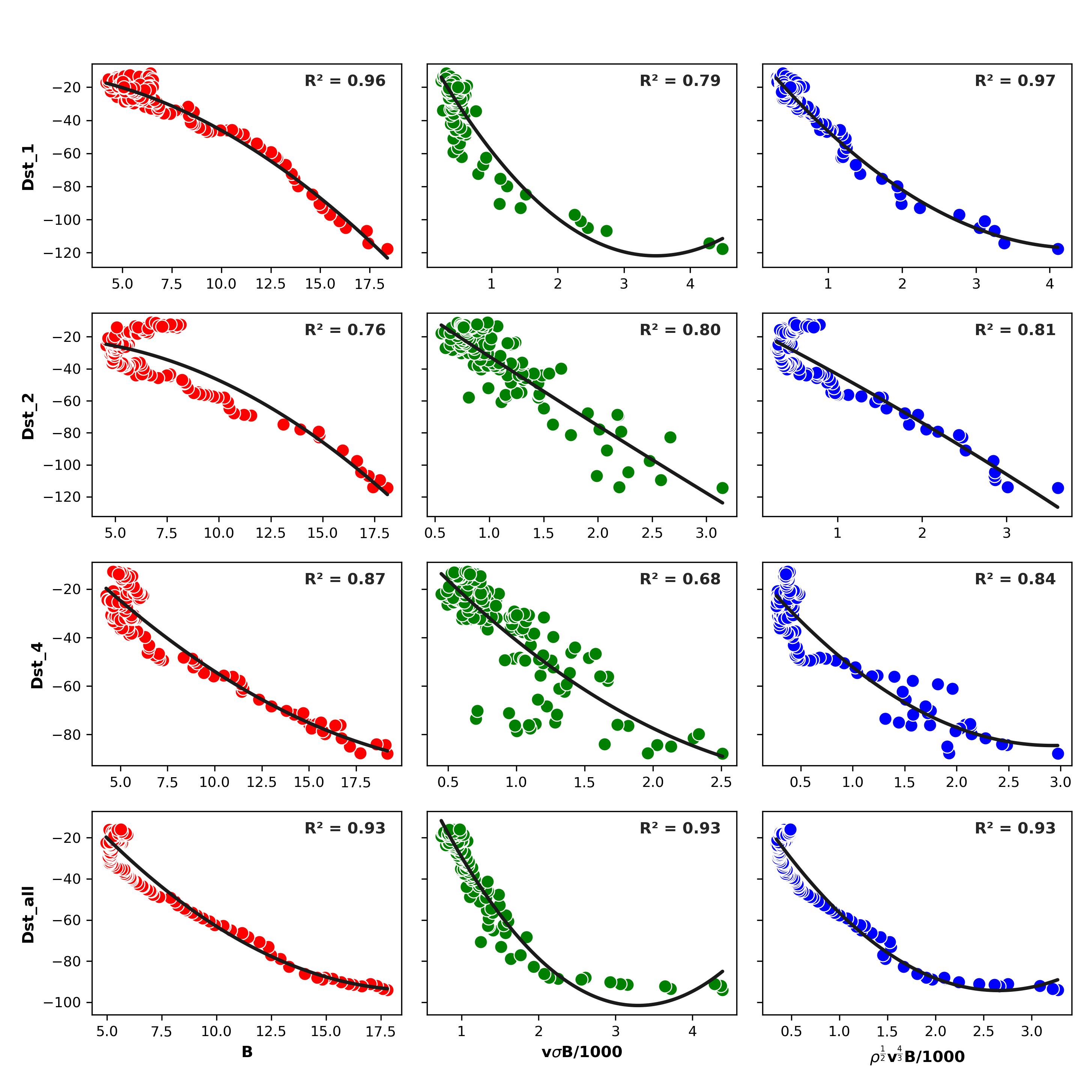}           
   \caption{The plots demonstrate the relation between Dst and some of IP SW parameters as well as derived functions based on recovery morphology of magnetic storms. 
The data points of the single, two, and three parameters/functions are represented by the filled circles in red, green, and blue, which correspond to the first, second, and third column panels, respectively. The black curved lines represent the most suitable parabolic curves for the data.
Four groups (very-fast, fast, very-slow, whole data set, CME, CIR, and slow) are represented by four rows (Dst-1, Dst-2, Dst-4, Dst-all). Each group comprises three classes (A, B, and C) that are represented as (red, green, and blue), respectively.\label{fig:d156}}
\end{figure}
\begin{figure}
\includegraphics[width=\textwidth]{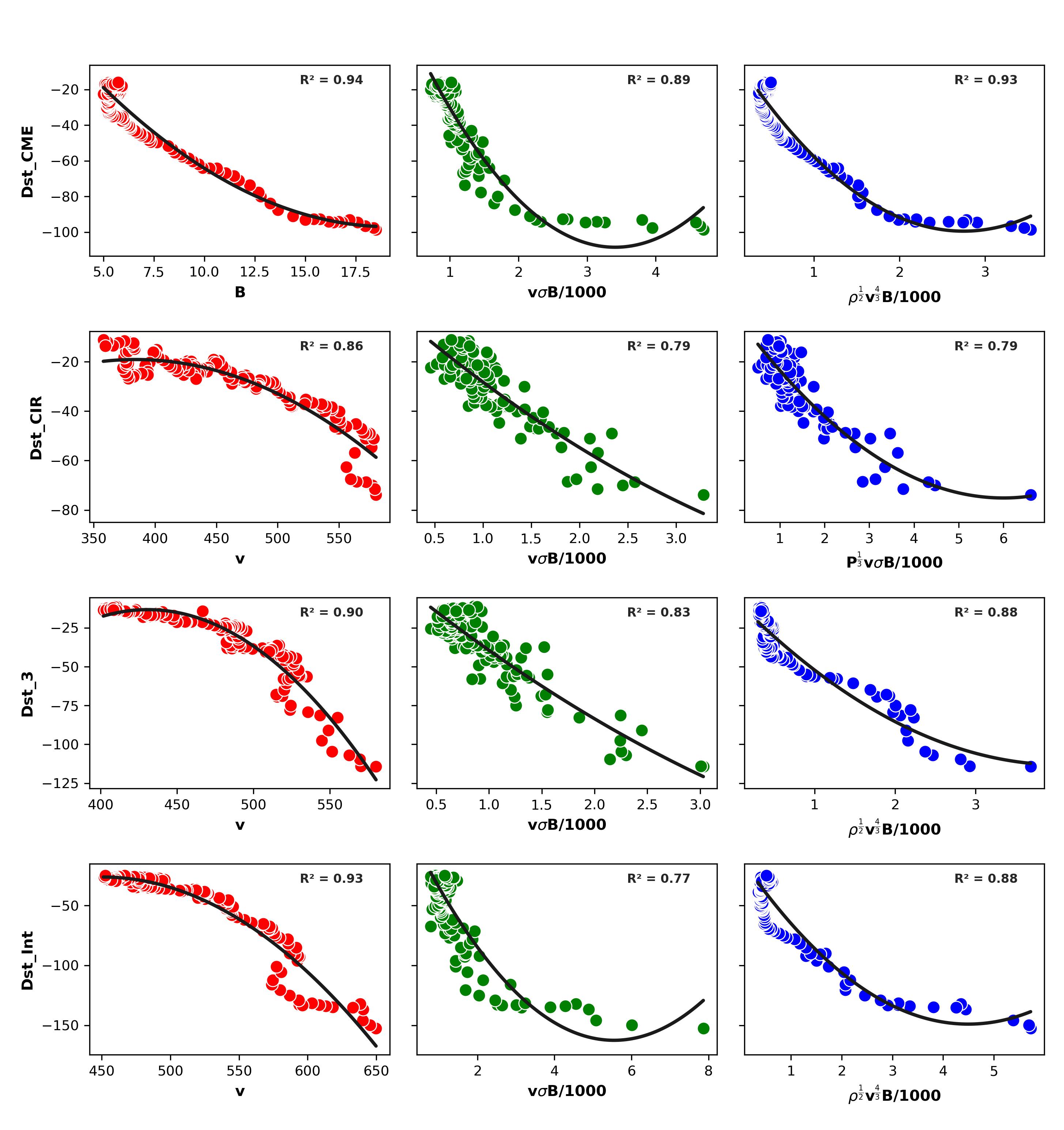}           
\caption{
The plots adopted from Figure \ref{fig:d156}, are based on four groups CME, CIR, slow and intense recovery are represented by four rows, Dst-CME, Dst-CIR, Dst-3 and Int respectively.
\label{fig:d157}}
\end{figure}
Based on the recovery morphology of GSs, the parabolic relationship between the Dst and IP SW parameters and derived functions has been analyzed from the result of superposed analysis. It has been determined that the recovery phase begins at the Dst minimum and lasts until the Dst reaches a quiet level. In order to accomplish this, we separately examined eight groups (r$^1$, r$^2$, r$^3$, r$^4$, entire data, CME, CIR and In as indicated in Table \ref{tabl1}).\\
Figure (\ref{fig:d156} \& \ref{fig:d157}) shows the scatter plots along with the second order polynomial fit between Dst and IP SW parameters and derived CFs, which were selected as the best correlated IP SW field and plasma parameters/functions. Best IP parameter/function from each class (A, B \& C) as (single, double \& triple) have represented as (red, green \& blue filled circles) respectively, in Figures (\ref{fig:d156} \& \ref{fig:d157}). The black curved lines are the best fitting parabolic curves.
We have fitted our data using the second order polynomial Y(X) = AX$^2$+BX+C. 
Figures (\ref{fig:d156} \& \ref{fig:d157}) shows a total of eight rows (Dst-1, Dst-2, Dst-4, Dst-all, Dst-CME, Dst-CIR, Dst-3 \& Int) corresponding to recovery groups (very-fast, fast, very-slow, whole data set, CME, CIR, slow and intense), respectively.
The three columns (red, green, and blue) represent (single parameter, two parameter CFs, and three parameter CFs) respectively, for each of the eight recovery groups.\\
Generally, IP SW parameters (IMF B, speed v and IMF Sigma vector $\sigma$B) and derived CFs (v$\sigma$B/1000, $\rho^{\frac{1}{2}}$v$^{\frac{4}{3}}$B/1000) play significant role to control the recovery dynamics of geomagnetic disturbance. The combination of function v$\sigma$B/1000 with viscus term $\rho$ \& pressure P with power indices of (1, $\frac{1}{2}$, $\frac{1}{3}$) are also well representing the recovery dynamics of magnetic storms, as their coefficient of determination R$^2$ is tabulated in Table \ref{ttx}.
The total IMF B correlates best for groups (r$^1$, r$^2$, r$^4$, entire data set, CME) and SW speed v best correlates for groups (CIR, r$^3$ and Int) as a single parameter class$-$A. The two parameter derived CF v$\sigma$B/1000 shows the best correlation with recovery dynamics of GSs for class$-$B for all groups. 
The three parameter CFs $\rho^{\frac{1}{2}}$v$^{\frac{4}{3}}$B/1000 have the best correlation with storm recovery morphology for class$-$C, except for the CIR group, which is significantly correlated with P$^{\frac{1}{3}}$v$\sigma$B/1000, implying that CIR-driven GSs are highly turbulent show high fluctuation during recovery phase.
Our findings support previous research on the south-north IMF Bz fluctuations of CIR-triggered storms throughout the recovery phase \citep[e.g.][]{2006ilws.conf..266G, 2009JASTP..71..885G}. \\
The best fit equations for the two single parameters (B \& v) are $\mathrm{Dst(B)=-0.30B^2-0.64B-9.51}$ and Dst(v) $=-$0.00v$^2$$+$3.20v$-$746.24, respectively. During the very-fast recovery group, the IMF B has the strongest relationship with Dst, with a coefficient of determination of R$^2=0.96$. Meanwhile, the SW velocity v has the strongest correlation with the Intense group, with a return of coefficient of determination R$^2=$ 0.93.
Two parameter CF v$\sigma$B/1000 has the strongest relationship with Dst for all groups, especially, for the entire data set group with coefficient of determination R$^2=0.93$, with best fit equation $\mathrm{Dst(v\sigma B/1000)=13.73(v\sigma B/1000)^2-90.42(v\sigma B/1000)}$\\$+47.29$.
$\mathrm{Dst(\rho^{\frac{1}{2}}v^{\frac{4}{3}}B/1000)=6.09(\rho^{\frac{1}{2}}v^{\frac{4}{3}}B/1000)^2-53.68(\rho^{\frac{1}{2}}v^{\frac{4}{3}}B/1000)+1.01}$ and\\ $\mathrm{Dst(P^{\frac{1}{3}}v\sigma B/1000)=2.06(P^{\frac{1}{3}}v\sigma B/1000)^2-24.76(P^{\frac{1}{3}}v\sigma B/1000)-0.83}$ are the best fit equations for the three-parameter SW-magnetosphere electrodynamical CFs ($\rho^{\frac{1}{2}}$v$^{\frac{4}{3}}$B/1000 \& P$^{\frac{1}{3}}$v$\sigma$B/1000), respectively. With a coefficient of determination of R$^2=0.97$, the $\rho^{\frac{1}{2}}$v$^{\frac{4}{3}}$B/1000 exhibits the highest correlation with Dst during the very-fast recovery group. In the meantime, the CIR recovery group and the perturbation-related function P$^{\frac{1}{3}}$v$\sigma$B/1000 have a strong relationship, with a return of coefficient of determination R$^2=0.79$.\\
Overall, the three-parameter SW-magnetosphere electrodynamical CF, which combines the viscus term ($\rho^{\frac{1}{2}}$) and the electric field-related function (v$^{\frac{4}{3}}$B/1000) ($\rho^{\frac{1}{2}}$v$^{\frac{4}{3}}$B/1000), significantly impacts the recovery phase morphology of geomagnetic disturbances.

\begin{table}
\caption{Presents the results of correlation obtained from second order polynomial fits conducted between the Dst and IP SW parameters, along with their products for eight groups. The coefficient of determination R$^2$ were determined based on the recovery profiles or morphologies. 
}\label{ttx}
\begin{tabular*}{\textwidth}{@{\extracolsep\fill}lccccccccc}
\toprule%
No&\textbf{Param.} & \multicolumn{8}{@{}c@{}}{\textbf{Groups}} \\ \cmidrule{3-10}%
& & \textbf{all} & \textbf{r$^1$} & \textbf{r$^2$}&\textbf{r$^3$} & \textbf{r$^4$}& \textbf{CME}& \textbf{CIR} & \textbf{Int}\\
\midrule
1&B &0.921$^{\beta}$ &0.96&0.76&0.85&0.87&0.94&0.73&0.92\\
2&Bz &0.35 &0.57&0.61&0.50&0.53&0.45&0.24&0.34\\
3&$\sigma$B &0.90 &0.63&0.70&0.79&0.74&0.86&0.65&0.78\\
4&$\rho$ &0.57 &0.41&0.52&0.58&0.63&0.53&0.68&0.53\\ 
5&v &0.85 &0.93&0.67&0.90&0.83&0.90&0.86&0.925$^{\beta}$\\
6&P &0.84 &0.78&0.61&0.80&0.75&0.82&0.65&0.81\\
7&v$\sigma$B/1000 &0.924$^{\beta}$ &0.79&0.80&0.83&0.68&0.89&0.79&0.77\\
8&$\rho$v$\sigma$B/1000 &0.81 &0.69&0.68&0.77&0.79&0.77&0.69&0.75\\
9&$\rho^{\frac{1}{2}}$v$\sigma$B/1000 &0.86&0.73&0.74&0.83&0.83&0.82&0.71&0.79\\
10&$\rho^{\frac{1}{3}}$v$\sigma$B/1000 &0.88&0.75&0.76&0.84&0.82&0.84&0.71&0.81\\
11&Pv$\sigma$B/1000 &0.75 &0.67&0.78&0.82&0.80&0.71&0.73&0.64\\
12&P$^{\frac{1}{2}}$v$\sigma$B/1000 &0.86  &0.74&0.80&0.87&0.80&0.82&0.79&0.73\\
13&P$^{\frac{1}{3}}$v$\sigma$B/1000 &0.86  &0.75&0.81&0.87&0.82&0.85&0.80&0.75\\
14&Pv$^{\frac{4}{3}}$Bz/1000 &0.30  &0.29&0.22&0.45&0.09&0.26&0.16&0.31\\
15&P$^{\frac{1}{2}}$v$^{\frac{4}{3}}$Bz/1000 &0.37  &0.39&0.55&0.40&0.19&0.33&0.32&0.32\\
16&P$^{\frac{1}{3}}$v$^{\frac{4}{3}}$Bz/1000 &0.37  &0.39&0.63&0.38&0.24&0.34&0.32&0.29\\
17&$\rho^{\frac{1}{2}}$v$^{\frac{4}{3}}$B/1000 &0.927$^{\beta}$  &0.97&0.81&0.88&0.84&0.93&0.66&0.88\\ 
18&$\rho^{\frac{1}{3}}$v$^{\frac{4}{3}}$Bz/1000 &0.43  &0.44&0.69&0.48&0.35&0.40&0.31&0.37\\
19&$\rho^{\frac{1}{2}}$v$^{\frac{4}{3}}$Bz/1000 &0.44  &0.45&0.71&0.51&0.34&0.40&0.32&0.41\\
20&$\rho$v$^{\frac{4}{3}}$Bz/1000 &0.42  &0.33&0.67&0.58&0.09&0.36&0.43&0.41\\
21&Ey &0.40  &0.51&0.57&0.43&0.38&0.37&0.26&0.22\\
22&$\rho$Ey &0.45  &0.36&0.70&0.58&0.09&0.38&0.44&0.42\\
23&$\rho^{\frac{1}{2}}$Ey &0.46  &0.47&0.72&0.53&0.40&0.42&0.32&0.42\\
24&$\rho^{\frac{1}{3}}$Ey &0.44 &0.51&0.70&0.51&0.40&0.42&0.30&0.39\\
25&PEy &0.31  &0.28&0.25&0.48&0.10&0.27&0.16&0.32\\
26&P$^{\frac{1}{2}}$Ey &0.39  &0.41&0.60&0.43&0.23&0.35&0.32&0.34\\
27&P$^{\frac{1}{3}}$Ey &0.40  &0.42&0.66&0.42&0.29&0.37&0.31&0.32\\
28&v$^{\frac{4}{3}}$Bz/1000 &0.35&0.50&0.66&0.38&0.33&0.33&0.26&0.19\\
\botrule
\end{tabular*}
\footnotetext{To obtain the ideal parameter, $^{\beta}$ indicate the values to the third decimal place.}
\end{table}
\section{Summary and Conclusions}\label{sec:conc}
In the present study, geomagnetic index Dst, IP SW plasma and field parameters, and derived CFs for 1-hour averaged resolution have been retrieved. The study focused on analyzing the recovery characteristics of GSs by examining 57 events with Dst $\leq50$ nT. Most of the analyzed events are triggered by CMEs and small amount of events driven by CIRs.
For both storms driven by CMEs and CIRs, we fitted exponential curves during the recovery phase. 
We may conclude from these results that the ring current decays faster for storms with short main phase durations in general, and it decays quickly for one day followed by gradual long-lasting second-step recovery. Magnetic storms with longer main phase duration, on the other hand, exhibit one step slow and gradual recovery.
Storms with shorter main and recovery phases show two-step recovery where as storms long lasting main and recovery durations show one step sluggish recovery. Only, intense geomagnetic disturbances show two-step recovery. During recovery of very intense storms (Dst $\leq-150$ nT) only the first initial phase (28\%) of recovery period has perturbed. Notably, intense storms fluctuates for short time and moderate storms continue their recovery with turbulence for longer periods.\\
We used linear correlation analysis for the recorded recovery time constants between Dst and IP SW parameters/functions to get the best parameter which represent the recovery time of magnetic storms. The decay rate of geomagnetic disturbances, as observed through the Dst index, could be best described by the decay rates of the function ($\rho^{\frac{1}{2}}$Ey) with the corresponding Pearson correlation coefficient 0.791, where both viscus term ($\rho$) and the dawn-dusk electric field (Ey) contribute to determine the recovery rate of geomagnetic disturbances. Earlier works suggested 
that the ring current decay is best described by electric field related function [vBs (-vBz) or Ey].
Our analysis suggests that, this electric field related function when coupled with a viscus function ($\rho^{\frac{1}{2}}$) (i.e, $\rho^{\frac{1}{2}}$Ey) better describe the ring current decay compared to electric field related function Ey (-Bz) alone.\\
The recovery phase morphology demonstrates a highly significant dynamic relationship between Dst and IP SW parameters/functions. IP SW parameters (IMF B, speed v and IMF Sigma vector $\sigma$B) and derived CFs (v$\sigma$B, $\rho^{\frac{1}{2}}$v$^{\frac{4}{3}}$B) plays significant role to control the recovery dynamics of geomagnetic disturbance.
The recovery morphology of geomagnetic disturbances is strongly correlated with (IMF B, CF v$\sigma$B, $\rho^{\frac{1}{2}}$v$^{\frac{4}{3}}$B) in Classes (A, B, \& C, see Section \ref{curve}) as (single, two parameter CF, three parameter CF), respectively.\\
These findings contribute to our understanding of how SW controls the ring current decay during geomagnetic disturbances. 
The decay rate of CF, which includes the viscus term $\rho$ and the electric field function ($\rho^{\frac{1}{2}}$Ey), appears to show a significant impact on the recovery rate of Dst, implying to be an important factor impacting the the ring current decay rate. 
On the other hand, the three-parameter SW-magnetosphere electrodynamical CF, which combines the viscus term ($\rho^{\frac{1}{2}}$) and the electric field-related function (v$^{\frac{4}{3}}$B/1000) ($\rho^{\frac{1}{2}}$v$^{\frac{4}{3}}$B/1000), significantly impacts the recovery phase morphology of geomagnetic disturbances.
Overall, this study enhances our knowledge of the recovery dynamics of magnetic storms and their relationship with various geomagnetic and interplanetary parameters, providing valuable insights into the complex processes involved in geomagnetic storm recovery.
Since the recovery phase of GS is nonlinear, and recovery process is complex, this work instigates and contributes towards this concern. However, it is desirable to continue the efforts towards better understanding of the mechanisms and processes during recovery of geomagnetic disturbances.

\bmhead{Author contributions}
OA planned the study, wrote manuscript, and participated in the data analysis. BB contributed to design the approach, review and editing. MD took part in the interpretation, review and editing process.
\bmhead{Data Availability}
Analyzed data were extracted from NASA/OMNIWeb, available on \url{http://omniweb.gsfc.nasa.gov.}
\bmhead{Competing interests}
The authors declare no competing interests.
\bmhead{Acknowledgements}

This study used data obtained from the GSFC/SPDF OMNIWeb interface, which can be found at \url{https://omniweb.gsfc.nasa.gov}. We would like to thank the reviewers for their insightful feedback and constructive suggestions, which considerably improved the quality of this work.

\bibliography{spr}

\end{document}